\documentclass{aa}
\usepackage{graphicx} 
\usepackage{txfonts}
\usepackage[colorlinks = true,
            linkcolor = blue,
            urlcolor  = black,
            citecolor = cyan]{hyperref}

\defcitealias{deBlok+24}{dB24}

\newcommand{\hi}{H\,\textsc{i}}

\newcommand {\kms} {\,{\rm km\,s}^{-1}}

\newcommand {\pc} {\,{\rm pc}}

\newcommand {\kpc} {\,{\rm kpc}}
\newcommand {\Mpc} {\,{\rm Mpc}}
\newcommand {\cmmq}{\,{\rm cm^{-2}}}
\newcommand {\cmmc}{\,{\rm cm^{-3}}}

\newcommand {\miJyb} {\,{\rm mJy\,beam}^{-1}}

\newcommand {\de}{^{\circ}}

\newcommand {\msun}{\,{\rm M}_\odot}

\newcommand{\zsun}{\,{Z}_\odot}

\newcommand{\Myr}{\,{\rm Myr}}

\newcommand{\K}{\,{\rm K}}

\newcommand {\msunyr}{\,{{\rm M}_\odot\,\rm yr}^{-1}}

\newcommand{\avg}[1]{\left< #1 \right>} 

\def\apjl{ApJ}%
\def\apjs{ApJS}%
\def\ao{Appl.~Opt.}%
\def\aap{A\&A}%
\defcitealias{Posti+19}{PFM19}

\begin{document}

\title{HI within and around observed and simulated galaxy discs}
\subtitle{Comparing MeerKAT observations with mock data from TNG50 and FIRE-2}
\titlerunning{HI around observed and simulated galaxies}
\authorrunning{A. Marasco et al.}

\author{A. Marasco\inst{1},
          W.\,J.\,G. de Blok\inst{2,3,4},
          F.\,M. Maccagni\inst{5},
          F. Fraternali\inst{4},
          K.\,A. Oman\inst{6,7,8},
          T. Oosterloo\inst{2,4},
          F. Combes\inst{9},
          S.S. McGaugh\inst{10},
          P. Kamphuis\inst{11},
          K. Spekkens\inst{12},
          D. Kleiner\inst{2},
          S. Veronese\inst{2,4},
          P. Amram\inst{13},
          L. Chemin\inst{14},
          E. Brinks\inst{15}
          }
\institute{INAF – Padova Astronomical Observatory, Vicolo dell’Osservatorio 5, I-35122 Padova, Italy\\
\email{antonino.marasco@inaf.it}
\and
Netherlands Institute for Radio Astronomy (ASTRON), Oude Hoogeveensedijk 4, 7991 PD Dwingeloo, the Netherlands
\and
Dept.\ of Astronomy, Univ.\ of Cape Town, Private Bag X3, Rondebosch 7701, South Africa
\and
Kapteyn Astronomical Institute, University of Groningen, PO Box 800, 9700 AV Groningen, The Netherlands
\and
INAF – Osservatorio Astronomico di Cagliari, Via della Scienza 5, 09047 Selargius (CA), Italy
\and
Institute for Computational Cosmology, Durham University, South Road, Durham DH1 3LE, United Kingdom
\and
Centre for Extragalactic Astronomy, Durham University, South Road, Durham DH1 3LE, United Kingdom
\and
Department of Physics, Durham University, South Road, Durham DH1 3LE, United Kingdom
\and
Observatoire de Paris, LERMA, Collège de France, CNRS, PSL University, Sorbonne University, 75014 Paris, France
\and
Department of Astronomy, Case Western Reserve University, 10900 Euclid Avenue, Cleveland, OH, 44106, USA
\and
Ruhr University Bochum, Faculty of Physics and Astronomy, Astronomical Institute (AIRUB), 44780, Bochum, Germany
\and
Department of Physics, Engineering Physics and Astronomy, Queen's University, Kingston, ON, K7L 3N6, Canada
\and
Aix Marseille Univ, CNRS, CNES, LAM, Marseille, France
\and
Universidad Andr\'es Bello, Facultad de Ciencias Exactas, Departamento de Ciencias F\'isicas - Instituto de Astrof\'isica, Fernandez Concha 700, Las Condes, Santiago, Chile
\and
Centre for Astrophysics Research, University of Hertfordshire, College Lane, Hatfield, AL10 9AB, UK
}
   \date{Received ; accepted}

 
\abstract
{
Extragalactic gas accretion and outflows driven by stellar and AGN feedback are expected to influence the distribution and kinematics of gas in and around galaxies.
Atomic hydrogen (\hi) is an ideal tracer of these processes, and it is uniquely observable in nearby galaxies.
Here we make use of wide-field ($1\de\times1\de$), spatially resolved (down to $22''$), high-sensitivity ($\sim10^{18}\cmmq$) \hi\ observations of $5$ nearby spiral galaxies with stellar mass of $\sim5\times10^{10}\msun$, taken with the MeerKAT radio telescope. Four of these were observed as part of the MHONGOOSE survey. We characterise the main \hi\ properties in a few hundred kpc regions around the discs of these galaxies, and compare these with synthetic \hi\ data from a sample of $25$ similarly massive star-forming galaxies from the TNG50 (20) and FIRE-2 (5) suites of cosmological hydrodynamical simulations.
Globally, the simulated systems have \hi\ and molecular hydrogen (H$_2$) masses in good agreement with the observations, but only when the H$_2$ recipe of Blitz \& Rosolowsky (2006) is employed. The other recipes that we tested overestimate the H$_2$-to-\hi\ mass fraction by up to an order of magnitude.
On a local scale, we find two main discrepancies between observed and simulated data.
First, the simulated galaxies show a more irregular \hi\ morphology than the observed ones due to the presence of \hi\ with column density $<10^{20}\cmmq$ up to $\sim100\kpc$ from the galaxy centre, in spite of the fact that they inhabit more isolated environments than the observed targets.
Second, the simulated galaxies and in particular those from the FIRE-2 suite, feature more complex and overall broader \hi\ line profiles than the observed ones. 
We interpret this as being due to the combined effect of stellar feedback and gas accretion, which lead to a large-scale gas circulation that is more vigorous than in the observed galaxies. 
Our results indicate that, with respect to the simulations, gentler processes of gas inflows and outflows are at work in the nearby Universe, leading to more regular and less turbulent \hi\ discs.}

\maketitle
%

\section{Introduction} \label{s:intro}
The build-up of galaxies in the field environment can be thought of as the competition between the `positive' process of cooling and gravitational collapse of gas within the potential wells provided by dark matter (DM) halos \citep{WhiteRees78}, and a series of `negative' mechanisms such as gas heating and subsequent expulsions caused by feedback from star formation \citep{Larson74,DekelSilk86} and active galactic nuclei \citep[AGN;][]{SilkRees98,Harrison+17}.
This competition produces a continuous exchange of gas between the galaxy and its host halo, ultimately regulating the main properties (mass, angular momentum, chemical composition) of the interstellar medium (ISM), the material from which stars form and super-massive black holes grow.

The study of this galaxy-halo interplay is pivotal to shed light on a number of open issues in galaxy evolution theory.
Total gas consumption timescales in present-day galaxies are of order a few gigayears \citep{Kennicutt98,Leroy+13} and become even shorter at higher redshift \citep{Saintonge+13}, providing indirect but compelling evidence for the existence of a continuous gas supply that replaces the material used for star formation processes \citep{Fraternali&Tomassetti12}.
Cosmological simulations in a $\Lambda$ cold dark matter (CDM) framework identify such a supply in the form of cold ($10^5\K$) gas filaments either inflowing directly onto galaxies from the cosmic web (cold-mode accretion), or shock-heated to the halo virial temperature into a diffuse, extended gas reservoir, sometimes called `corona' \citep[hot-mode accretion,e.g.,][]{Keres05,Nelson+13,Ramesh+23a}.

However, direct evidence for gas accretion onto galaxies, at least in the local Universe where the most sensitive data are available, are scarce. 
Wet mergers can provide only a small fraction of the accretion rate required to balance the galaxy star formation rate \citep{Sancisi+08, DiTeodoroFraternali14}. 
The same applies for the supply provided by the direct infall of isolated, optically-dark \hi\ clouds detected in the halos of galaxy discs \citep{Kamphuis+22}.
Cosmological cold accretion is expected to produce relatively strong inwards radial motions at the periphery of discs \citep[e.g.][]{Trapp+22}, which seem to be elusive in observations \citep{DiTeodoroPeek21}.
While the expected radiative cooling rate of the hot corona is largely insufficient to match the star formation rate in galaxies like the Milky Way \citep[e.g.][]{MillerBregman15}, whether or not cold gas clouds can spontaneously form within the corona and precipitate onto the disc is a highly debated topic \citep{Binney+09,Nipoti10,Voit+17,Sobacchi+19,Afruni+23}.
Hence, an important point to address is how and where gas accretion onto galaxies takes place.
A possibility could be that the cooling of the coronal gas is limited to a region of a few kpc thickness surrounding the star formation disc \citep[e.g.][]{Stern+24}. 
The cooling of the coronal gas can be favoured by the interaction with material ejected from the galaxy by supernova feedback (galactic fountain), which produces a gas mixture with temperature, density and metallicity intermediate between those of the corona and of the ISM, thus with a short ($<100\Myr$) cooling time \citep{Marinacci+10,Fraternali17}.
This process is expected to leave a subtle, but detectable, imprint on the kinematics of the \hi\ at the disc-corona interface, characterised by a lagging rotation and a global inflow motion in the vertical and radial directions.
This can be exploited to infer the condensation rate of the coronal gas via the modelling of highly sensitive, spatially resolved \hi\ observations \citep{FB08, Marasco+12, Li+23}.

Galactic winds triggered by star formation or AGN feedback are thought to play a key role in setting two key properties of the observed galaxy population.
One is the low overall star formation efficiency: galaxies are unable to convert the majority of baryons associated with their DM halos into stars, with an efficiency that peaks in local $L_\star$ galaxies and rapidly declines for lower and higher luminosities \citep[e.g.][]{Moster+13, Behroozi+13}.
Thus most baryons in dark matter halos are not in the form of stars and the ISM, but in a more diffuse circumgalactic medium (CGM).
The other key property is the specific angular momentum ($j$) distribution (AMD) of stars and gas within galaxies.
Most star-forming galaxies across all redshifts appear as disc-like, rotationally-supported structures, and therefore are characterised by an AMD that lacks very low $j$ material.
Instead, typical dark matter halos in $\Lambda$CDM cosmological simulations have an excess of low angular momentum material compared to galaxy discs \citep[e.g.][]{Bullock+01}.
If dark matter and baryons in halos follow the same AMD \citep{SharmaSteinmetz05}, this mismatch implies a biased assembly, where galaxies have been forming their stars using preferentially the high $j$ reservoir available in their halo.
These two properties can be explained by powerful galactic winds that preferentially remove the low-$j$ star-forming material from the galaxy, lowering the star formation efficiency while promoting the build-up of galaxy discs.

Galaxy-scale outflows are routinely observed at all redshifts using a variety of different tracers \citep[e.g.][]{Veilleux+05,Rubin+14,Woo+16,Cicone+18}. 
However, robust measurements of the properties of the galactic wind such as speed and mass outflow rate have proved to be challenging due to the complex geometry and multi-phase nature of the phenomenon, often leading to different estimates depending on the method adopted \citep[e.g.][]{Chisholm+17,McQuinn+19,Concas+22,Marasco+23a}.
On the other hand, theoretical predictions for the wind properties vary depending on the model considered, with large-scale cosmological simulations in the $\Lambda$CDM framework typically predicting more `extreme' winds \citep[e.g.][]{Nelson+19b, Mitchell+20} compared to small-scale simulations of the ISM \citep[e.g.][]{KimOstriker18} or to chemical evolutionary models \citep[e.g.][]{CampsFarina+23, Kado-Fong+24}.

We can summarise the scenario described above as follows: first, there are strong theoretical arguments and compelling indirect evidence - but very little observational support - for the need for gas accretion onto galaxies; second, the feedback efficiency required by cosmological simulations is in tension with several observational constraints and small-scale simulations.
These tensions are caused, at least in part, by two limitations.
First, there exists a gap between hydrodynamical models of galaxy evolution and observations in terms of comparison strategy, as models were often used to derive physical quantities that could not readily be measured observationally.
The situation has drastically improved in the last few years \citep[e.g.][]{Marasco+18,Oman+19,Faucher+23,Gebek+23,Baes+24} thanks to the level of sophistication reached by present-day models, which permit the production of synthetic (broad-band or emission-line) observations, incorporating all of the limitations of the desired facilities such as sensitivity, resolution and field of view, providing a more direct and straightforward comparison with the data.
Second, there exists an observational limitation due to the intrinsically low-density nature of the material involved in the gas circulation processes, which demands very deep observations to make its analysis possible \citep[e.g.][]{Popping09}.
If we limit the discussion to the atomic hydrogen (\hi) phase, which will be the subject of our study, there is general agreement between different simulation suites that the large-scale filamentary structure of the gas accreting from the cosmic web can be revealed only by pushing the \hi\ column density ($N_{\rm HI}$) sensitivity below values of $\simeq10^{19}\cmmq$ \citep{vandeVoort+19,Ramesh+23a}.
Single-dish radio observations can easily reach such sensitivities and have shown that \hi\ with $N_{\rm HI}\!<\!10^{19}\cmmq$ makes up to $\sim2\%$ of the total \hi\ mass budget in and around nearby star-forming galaxies \citep[e.g.][]{Pingel+18}.
Clearly, radio-interferometric observations are key to infer the spatial distribution and kinematics of the low column density \hi, but the sensitivity required was hardly achievable until today.

This study is an attempt to overcome both of these limitations by combining ultra-deep, spatially resolved \hi\ observations of five nearby galaxies, similar to the Milky Way (MW) in terms of baryonic mass and SFR, taken with the MeerKAT radio telescope, with synthetic \hi\ data of similarly massive spirals extracted from two widely-studied suites of hydrodynamical simulations in the $\Lambda$CDM framework, Illustris TNG50 \citep{Pillepich+19, Nelson+19} and FIRE-2 \citep{Hopkins+18b}.
We focus on MW-like objects alone as we expect them to feature substantial ongoing gas inflows and outflows due to their relatively short gas consumption timescales \citep[in contrast with lower mass galaxies, e.g.][]{McGaugh+17}, an argument which is strongly supported by different hydrodynamical simulations of galaxy evolution in the MW-like mass regime \citep[e.g.][]{Fernandez+12, vandeVoort+19, Trapp+22}.
The use of two different simulation suites allows us not to be biased towards a particular model. 
The focus on the \hi\ emission line is justified by the fact that, even though this phase is not expected to dominate the mass budget of the CGM \citep[unless we limit our study to a region of a few kpc around the disc, see][]{Marasco+19b}, \hi\ observations with MeerKAT combine depth, resolution and field of view to provide a unique, broad picture of the disc-halo interplay.
Our goal is twofold: on the one hand the comparison will furnish insights into the physical interpretation of specific features detected in the data, on the other hand it will capture the main differences between the models and the data, which gives us clues on how the former can be refined.

This paper is structured as follows.
In Section\,\ref{s:method} we provide a description of the observed and simulated galaxy samples used, and of the approach adopted to build synthetic \hi\ datacubes (`mocks') from the models.
In Section\,\ref{s:results} we show our main results, highlighting the main differences between the data and the mocks in terms of column density and velocity dispersion distributions.
Section\,\ref{s:discussion} focuses on understanding the origins of the difference found in terms of environment, stellar feedback and gas accretion rates.
We summarise our results and present our conclusions in Section\,\ref{s:conclusions}.

\section{Galaxy samples}\label{s:method}
\begin{table*}
\caption{Main properties of the five galaxies observed with MeerKAT and analysed in this study.}
\label{t:sample_obs}
\centering
\begin{tabular}{lccccc}
\hline\hline
\noalign{\smallskip}
 HIPASS     & Name      &   $D$   &  $M_{\star}$     &   $M_{\rm HI}$    & SFR   \\
            &           &  [\Mpc] & [$10^{10}\msun$] & [$10^{10}\msun$] & [$\msunyr$] \\
 (1) & (2) & (3) & (4) & (5) & (6)\\  
\noalign{\smallskip}
\hline
\hline
\noalign{\smallskip}
J0052-31  &  NGC0289   & 21.5  &    2.69  &    2.24  & 1.58\\
J0335-24  &  NGC1371   & 22.7  &    4.27  &    0.93  & 0.26\\
J0445-59  &  NGC1672   & 19.4  &    4.90  &    1.95  & 5.89\\
J0419-54  &  NGC1566   & 17.7  &    4.79  &    1.20  & 4.07\\
\hline
\noalign{\smallskip}
J0342-47  &  NGC1433   & 18.6  &    7.94  &     0.27  & 12.59\\
\hline\hline
\noalign{\smallskip}
\multicolumn{6}{p{0.6\textwidth}}{\small {\bf Notes:} (1) HIPASS identification. (2) NGC name. (3) Distance, taken from \citet{Leroy+19} for J0052-31 and J0335-24, and from \citet{Anand+21} for J0445-59 and J0419-54. (4)-(6) Stellar masses, \hi\ masses and SFRs. Details are given in \citetalias{deBlok+24} for galaxies in the MHONGOOSE survey, and in \citet{Stuber+23} for J0342-47. With the exception of J0342-47, all galaxies are part of the MHONGOOSE survey \citepalias{deBlok+24}.}
\end{tabular}
\end{table*}

Our strategy is to focus on regular star forming galaxies with stellar masses comparable to those of the Milky Way, as for these systems we expect a more active disc-halo gas circulation due to their relatively low gas consumption timescales. 
Our selection criteria for the simulated galaxy samples are based on this strategy. 
On the observational side, our analysis is limited by the paucity of ultra-deep, spatially resolved HI observations for galaxies with the required properties. 
To our knowledge, the observed subsample used for our study is the best compromise that we can currently achieve in terms of HI data quality and required galaxy properties.

\subsection{The observed sample}\label{ss:mhongoose}
\begin{figure*}
\begin{center}
\includegraphics[width=1.0\textwidth]{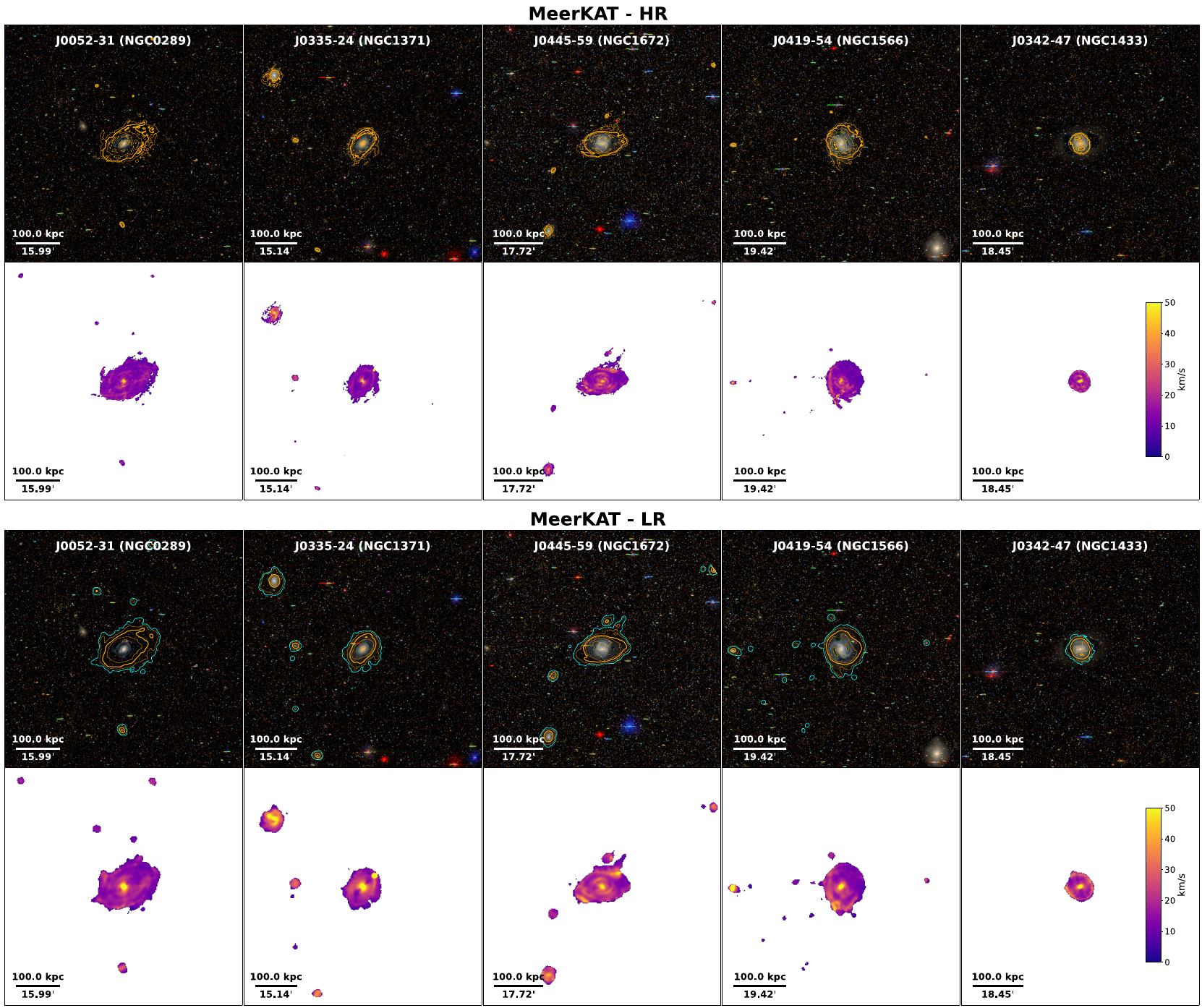}
\caption{The MeerKAT galaxy sample studied in this work.
The high ($\sim22"$) and low ($\sim65"$) resolution cases (HR and LR) are shown separately in the top and bottom parts of the figure.
\emph{Top panels:} \hi\ column density iso-contours overlaid on top of \emph{grz} colour-composite images from the DESI Legacy Imaging Surveys \citep{Dey+19}. Contours are drawn from primary-beam corrected moment-0 maps at N$_{\rm HI}$ levels of $10^{19}$, $10^{20}$ (thicker line), $10^{20.5}$ and $10^{21}\cmmq$. The LR panels show an additional contour at the level of $10^{18}\cmmq$, in cyan.
\emph{Bottom panels:} moment-2 maps. The velocity colour scheme is shown in the rightmost panels.}
\label{f:MHONGOOSE_momentmaps}
\end{center}
\end{figure*}

Our observed galaxy sample is formed by $5$ objects whose main properties are listed in Table \ref{t:sample_obs}.
With the exception of J0342-47, all galaxies have been observed as a part of the MeerKAT \hi\ Observations of Nearby Galactic Objects - Observing Southern Emitters (MHONGOOSE) survey \citep[hereafter \citetalias{deBlok+24}]{deBlok+24}.

 MHONGOOSE is a MeerKAT Large Survey Project aimed at providing ultra-deep, spatially resolved \hi\ observations of $30$ nearby ($D\lesssim20\Mpc$), galaxies outside clusters and major groups in the Southern hemisphere, covering a wide range of stellar and \hi\ masses ($6.0\!<\!\log_{10}(M_\star/\msun)\!<\!10.7$, $7.0\!<\!\log_{10}(M_{\rm HI}/\msun)\!<\!10.3$).
The goal of MHONGOOSE is to probe the \hi\ cycle between galaxy discs and their circumgalactic media and study its connection to star formation feeding and feedback processes.
To achieve this, MHONGOOSE exploits both the large FoV of MeerKAT, with a primary beam FWHM at $1.4$ GHz of $\sim1\de$ (corresponding to $350\kpc$ at a distance $D\!=\!20\Mpc$), and its outstanding multi-resolution performance, reaching a minimum \hi\ column density sensitivity ($3\sigma$ integrated over a velocity width of $16\kms$) of $6\times10^{17}\cmmq$ for a beam FWHM of $90"$ ($8.6\kpc$ at $D\!=\!20\Mpc$) for emission that is more extended than this resolution.
The sensitivity is better than $10^{20}\cmmq$ up to the highest resolution achievable, $7.6"$ ($0.74\kpc$ at $D\!=\!20\Mpc$).
The unique combination of area, sensitivity and resolution makes MHONGOOSE the optimal survey to study faint \hi\ features in and around galaxy discs.

As in this study we focus on galaxies with $M_\star$ similar to that of the MW, we are limited to the four most massive galaxies available in MHONGOOSE: J0335–24, J0052–31, J0419–54 and J0445–59 (see Table \ref{t:sample_obs} for the NGC identifications). 
These are the only MHONGOOSE galaxies with $M_\star\!>\!10^{10}\msun$ and have distances of $\sim20\Mpc$, which implies that the MeerKAT FoV covers approximately the virial region, assuming typical dark matter halos analogous to that of the MW \citep[e.g.][]{PostiHelmi19}.
Also, we limit our analysis to two of the six `standard' resolutions available in the survey, a high-resolution (HR) case, with beam FWHM of $\sim22"$ ($\sim2\kpc$) and nominal column density sensitivity ($3\sigma$ over $16\kms$) of $5\times10^{18}\cmmq$, and a low-resolution (LR) case with beam FWHM of $\sim65"$ ($6.3\kpc$) and sensitivity of $\sim10^{18}\cmmq$.
These two resolutions correspond, respectively, to the \texttt{r10$\_$t00} and \texttt{r05$\_$t60} configurations presented in \citetalias{deBlok+24}. 
Table \ref{t:obs_setup} lists the main datacube parameters associated with these two resolution cases, which will be adopted also for the realisation of synthetic \hi\ data for the simulated galaxy samples (see Section\,\ref{ss:building_mocks}).

\begin{table}
\caption{Characteristic parameters for the \hi\ datacubes of the observed and the simulated galaxy samples.}
\label{t:obs_setup}
\centering
\begin{tabular}{lccc}
\hline\hline
\noalign{\smallskip}
 Parameter &  unit &  HR &   LR   \\
\hline
\noalign{\smallskip}
Field of view $^a$ & $\de$ & $\sim1$ & $\sim1$ \\
                   & $\kpc$ & $\sim350$  & $\sim350$\\
Beam size $^b$ & $''$ & $26.4\times18.2$ & $65.3\times64.0$  \\
               & $\kpc$ & $2.6\times1.8$ & $6.3\times6.2$  \\
Beam PA & $\de$ &  $136$  & $92$ \\
Pixel size & $''$ & $5$ & $20$  \\
           & $\kpc$ & $0.5$ & $1.9$  \\
Velocity range & $\kms$ & $\pm500$ & $\pm500$\\
Channel width $^c$  & $\kms$ & $1.4$ & $1.4$  \\
$\sigma_{\rm rms}$ per channel & $\miJyb$ & $0.15$ & $0.25$  \\
Minimum $N_{\rm HI}$$^d$ & $\cmmq$ & $5\times10^{18}$ & $1\times10^{18}$  \\
\hline\hline
\noalign{\smallskip}
\multicolumn{4}{p{0.48\textwidth}}{\small {\bf Notes:} $^a$ corresponding to the FWHM of the MeerKAT primary beam. $^b$  Sizes in $\kpc$ are based on a distance of $20\Mpc$. $^c$ J0342-47 cubes have a different channel width of $5.5\kms$. $^d$ $3\sigma$ sensitivity integrated over $16\kms$. With respect to the values listed here, the rms noise per channel and the minimum detectable $N_{\rm HI}$ for J0342-47 are larger by $18$ per cent (see text).}
\end{tabular}
\end{table}

In addition to these four MHONGOOSE systems, we have added a fifth MW-like galaxy, J0342-47 (or NGC\,1433). 
This is a Seyfert galaxy that has stellar mass and distance compatible with the other galaxies in our sample (see Table \ref{t:sample_obs}), and is characterised by a central outflow visible in both the molecular \citep{Combes+13} and the ionised \citep{Arribas+14} phase driven by its AGN.
J0342-47 has been observed for $10$ hours with MeerKAT as a part of the MKT-20202 project (PI: F. Maccagni).
A channel width of $5.51\kms$ has been used for the data reduction, providing a column density limit about $18$ per cent higher than those reported in our Table \ref{t:obs_setup} for the same resolution cases.

Fig.\,\ref{f:MHONGOOSE_momentmaps} shows the primary-beam corrected moment-0 (column density, $N_{\rm HI}$) and moment-2 (velocity dispersion, $\sigma$)\footnote{In this study we use terms such as `moment-2 value', `velocity dispersion' and `$\sigma$' interchangeably, but we remind the reader that this may not hold in all contexts. For single (Gaussian) profiles, sigma and moment-2 are equivalent and associated with the line broadening due to kinetic temperature and turbulence effects. This is no longer the case for more complex profiles, such as those produced by multiple components (see Section \ref{ss:line_profiles}).}
\hi\ maps for the five MeerKAT galaxies and for both the resolution cases studied, together with \emph{gri} colour-composite images from the data release 10 of the DESI Legacy Imaging Surveys \citep{Dey+19}.
The \hi\ maps were derived using the SoFia-2 software package \citep{Westmeier+21} as discussed in \citetalias{deBlok+24}. 
We refer to that paper for more details; here we describe the basic procedure.
The SoFiA-2 `smooth and clip (S+C)' method was used to find significant voxels in the cubes using spatial kernels of $0$ and $4$ pixels and velocity kernels of $0$, $9$ and $25$ channels using a threshold of $4\sigma_{\rm rms}$. 
Detected voxel regions were then linked over a maximum of $5$ pixels spatially and $8$ channels spectrally ($2$ channels in J0342-47).
A minimum size of $4$ spatial pixels and $15$ channels was then imposed on these linked regions. 
The SoFiA reliability parameter was used to further identify significant sources in the cube. 
However, as mentioned in \citetalias{deBlok+24}, the value adopted for this parameter was not crucial as almost all sources were well separated from the noise, thanks to the high S/N of the MHONGOOSE data. 
The significant sources were extracted using a reliability value of $0.8$ and requiring an integrated S/N value of $5$.

The detailed \hi\ morphology and kinematics of these five galaxies will be investigated elsewhere. 
Here, we remark on a striking feature of these observations, which is the lack of diffuse \hi\ emission around the galaxies. Virtually all of the \hi\ visible around the main targets is in the form of \hi-bearing satellites.
In addition, all galaxies feature relatively low $\sigma$ values ranging from $8$ to $30\kms$, with the exception of their central region, where $\sigma$ increases due to beam-smearing effects and to the probable presence of outflows driven by feedback from an Active Galactic Nucleus (AGN), as the majority of the systems in the observed sample are known Seyfert galaxies.
The amount of \hi\ in the CGM and the \hi\ velocity dispersion are the two main features that will be contrasted with predictions from the simulations in this study.

\subsection{The simulated sample}\label{ss:simulations}
\begin{table}
\caption{Main properties of the simulated systems studied in this work.}
\label{t:sample_sims}
\centering
\begin{tabular}{lcccc}
\hline\hline
\noalign{\smallskip}
 ID & $M_{200}$      &   $M_\star$      & $M_{\rm HI}$    & SFR        \\
          & [$10^{12}\msun$] & [$10^{10}\msun$] & [$10^{10}\msun$] & [$\msunyr$] \\
(1) & (2) & (3) & (4) & (5)\\
\noalign{\smallskip}
\hline
\hline
\noalign{\smallskip}
\multicolumn{5}{c}{TNG50}\\
\hline
459557  &      0.81  &      4.50  &     1.67  &      1.34\\
476266  &      1.45  &      5.02  &     2.27  &      6.93\\
477328  &      1.95  &      6.22  &     3.32  &      8.91\\
517271  &      1.41  &      6.44  &     1.21  &      2.06\\
520885  &      1.17  &      5.47  &     3.17  &      2.90\\
522530  &      1.29  &      6.89  &     1.63  &      1.55\\
526478  &      1.12  &      4.84  &     2.60  &      7.40\\
531320  &      1.15  &      3.98  &     2.31  &      2.29\\
537488  &      0.95  &      4.38  &     0.89  &      3.15\\
537941  &      0.95  &      5.24  &     1.85  &      4.31\\
543376  &      0.93  &      6.14  &     1.11  &      4.14\\
543729  &      0.95  &      4.93  &     1.70  &      6.05\\
550934  &      0.89  &      3.56  &     1.77  &      4.72\\
554798  &      0.87  &      5.33  &     0.74  &      1.23\\
555013  &      0.89  &      3.66  &     1.85  &      1.46\\
555287  &      0.78  &      3.85  &     1.44  &      4.56\\
555601  &      0.78  &      4.60  &     1.75  &      4.49\\
559036  &      0.72  &      3.40  &     2.10  &      3.23\\
565251  &      0.71  &      3.55  &     0.62  &      3.60\\
566365  &      0.68  &      3.25  &     0.44  &      3.36\\ 
\hline
\noalign{\smallskip}
\multicolumn{5}{c}{FIRE-2}\\
\hline
m12b  &      1.43  &      9.93  &     1.13  &      9.04\\
m12f  &      1.71  &      9.59  &     2.70  &      10.94\\
m12i  &      1.18  &      7.26  &     1.27  &      6.97\\
m12r  &      1.10  &      2.06  &     1.13  &      2.39\\
m12w  &      1.08  &      6.52  &     0.25  &      10.79\\
\hline\hline
\noalign{\smallskip}
\multicolumn{5}{p{0.47\textwidth}}{\small {\bf Notes:} (1) Simulated galaxy ID. For TNG50 systems, this refers to the subhalo index in the group catalogues as determined by the \textsc{subfind} algorithm \citep{Dolag+09}. For FIRE-2 systems, it refers to the run name in \citet{Wetzel+23}. (2) Halo mass, computed within the radius where the mean halo density becomes equal to $200$ times the critical density of the Universe. (3)-(5) Stellar mass, \hi\ mass and instantaneous SFR computed within $100\kpc$ from the centre of each galaxy.}
\end{tabular}
\end{table}

\begin{figure}
\begin{center}
\includegraphics[width=0.49\textwidth]{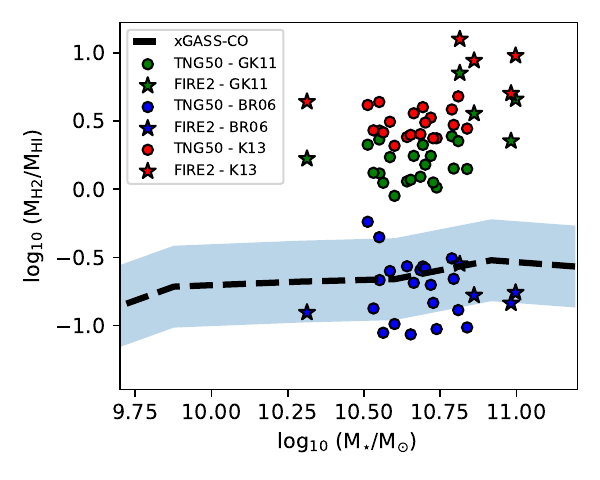}
\caption{Molecular-to-atomic hydrogen mass ratio as a function of galaxy stellar mass for the TNG50 (circles) and FIRE-2 (stars) systems studied in this work. All masses are computed within $100\kpc$ from the galaxy centre. Different H$_2$/\hi\ partition recipes are shown with different colours: green for \citet[][GK11]{GnedinKravtsov11}, red for \citet[][K13]{Krumholz13}, and blue for \citet[][BR06]{BlitzRosolowsky06}. The recipe of BR06 is the one that better reproduces the observational values from the xGASS Survey \citep{Catinella+18}, which are represented here by the shaded region.}
\label{f:MH2_over_MHI}
\end{center}
\end{figure}

\begin{figure*}
\begin{center}
\includegraphics[width=\textwidth]{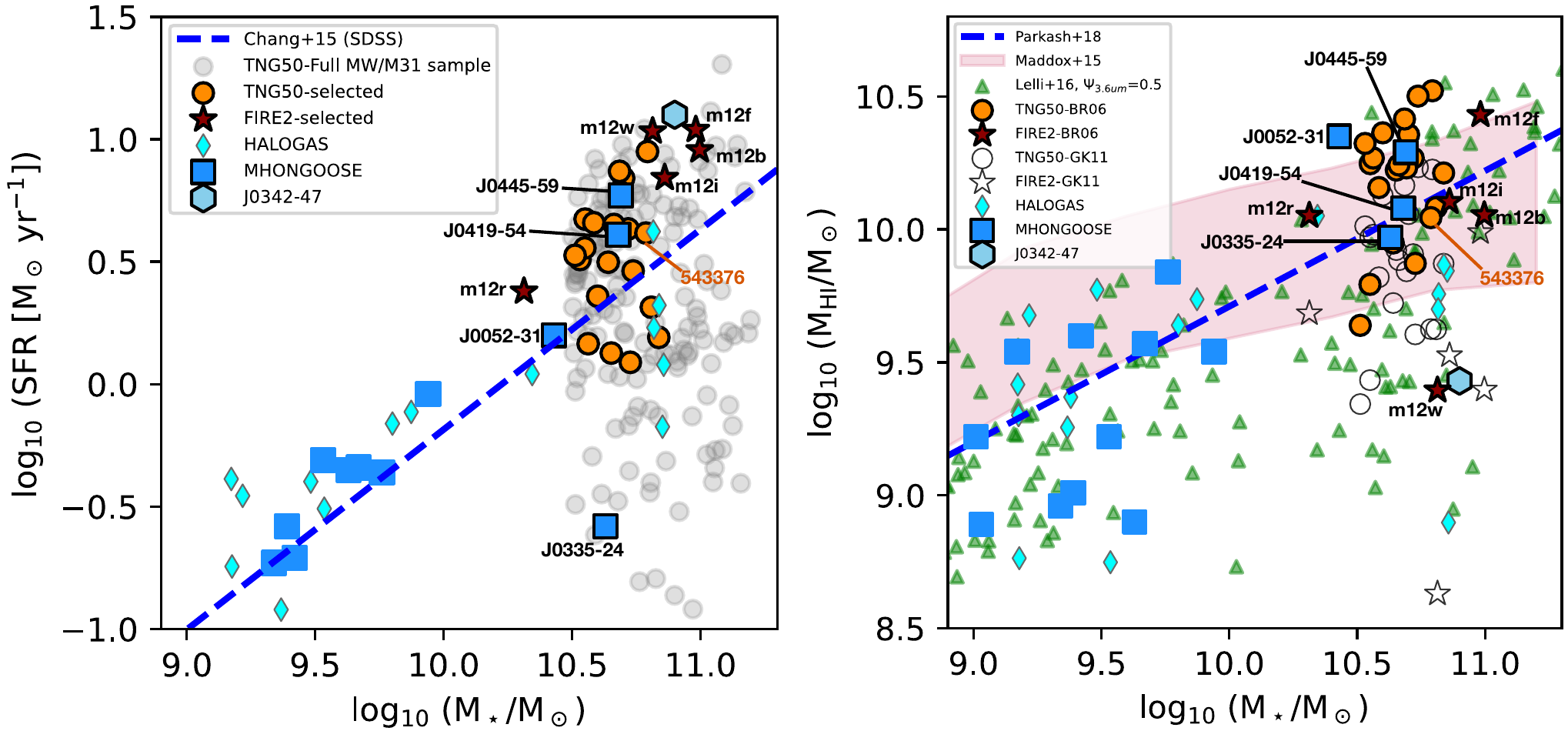}
\caption{Main scaling properties of the simulated galaxy samples compared to various observational datasets. \emph{Left panel:} SFR vs. $M_\star$ for MHONGOOSE \citepalias[][blue squares]{deBlok+24}, J0342-47 (light-blue hexagon), HALOGAS \citep[][cyan diamonds]{Heald+11,Kamphuis+22}, and our TNG50 (orange circles) and FIRE-2 (dark red stars) samples. The blue dashed line shows the SDSS main sequence from \citet{Chang+15}. Grey circles show the other systems from the MW/M31 TNG50 sample of \citet{Pillepich+24}. \emph{Right panel:}  $M_{\rm HI}$ vs. $M_\star$. Markers are as in the left panel, with the addition of the SPARC sample \citep[][green triangles]{Lelli+16}. 
The purple-shaded area shows the $\pm1\sigma$ confidence region around the median scaling found by \citet{Maddox+15}, while the blue dashed line shows the linear fit to the \hi-WISE data from \citet{Parkash+18}.
Masses for the simulated galaxies are computed within $100\kpc$ from their centre. Black-edged markers are used for systems that are studied in this work. For comparison, we also show the simulated galaxies with \hi\ masses computed with the recipe of \citet{GnedinKravtsov11} (empty stars and circles).}
\label{f:obs_vs_sim_global}
\end{center}
\end{figure*}

We focus on two different state-of-the-art simulation suites: Illustris TNG50 \citep{Pillepich+19, Nelson+19b} and FIRE-2 \citep{Hopkins+18}.
TNG50 is a cosmological, magneto-hydrodynamical simulation of galaxy formation and evolution in a $\Lambda$CDM universe \citep{Plank16}, made using the moving-mesh code AREPO \citep{Springel10,Weinberger+20} and adopting the fiducial Illustris TNG galaxy formation model \citep{Weinberger+17,Pillepich+18}.
It follows the evolution of dark matter, gas, stars, supermassive black holes (SMBHs) and magnetic fields in a $\sim(50\Mpc)^3$ cubic (co-moving) volume down to $z\!=\!0$, with a maximum resolution\footnote{expressed in terms of the minimum comoving value of the adaptive gas gravitational softening} of $\sim70\pc$ and gas particle mass of $8.5\times10^4\msun$. 
Amongst the different astrophysical processes modelled, TNG50 includes primordial and metal-line cooling down to $10^4\K$, heating from a spatially homogeneous UV/X-ray background, galactic-scale stellar-driven outflows, and the seeding, growth and feedback of SMBHs.
The cold and dense phase of the ISM is not directly resolved in TNG50, and star-forming gas is treated following the prescriptions of \citet{SpringelHernquist03} which employ a two-phase, effective equation-of-state model.

In this study, we focus on galaxies from the MW- and Andromeda-like (MW/M31-like) TNG50 sample recently presented by \citet[][see also \citealt{Ramesh+23a}]{Pillepich+24}. This sample is made of 198 grand design spiral galaxies, with $M_\star$ in the range $10^{10.5-11.2}\msun$, and within a MW-like large-scale environment at $z\!=\!0$.
The latter condition ensures the absence of massive ($M_\star\!>\!10^{10.5}$) companions within $500\kpc$ from the main galaxy, and having host halo masses below those of typical galaxy groups ($M_{200}<10^{13}\msun$), which qualitatively matches the environment of galaxies in MHONGOOSE.
From this sample we randomly extract 20 systems with a SFR that falls within $\pm0.5$ dex of the SFR-$M_\star$ relation of \citet{Chang+15}, to ensure star formation properties similar to those of similarly massive galaxies in the local Universe.
Additionally, we have applied a further cut to ensure a minimum \hi\ mass of $10^{9.5}\msun$, similar to that of J0342-47, the most \hi-poor galaxy of our observed sample \citep[$\log_{10}(M_{\rm HI}/\msun)=9.43\pm0.08$ assuming a distance of $18.63\pm1.86\Mpc$, from][]{Anand+21}. Details on the computation of \hi\ masses for the simulated galaxies are provided below.
The main properties of these 20 objects are listed in Table \ref{t:sample_sims}.

The FIRE-2 suite is a collection of zoom-in hydrodynamical simulations performed with the mesh-less finite-mass (MFM) method implemented in the \textsc{GIZMO} \citep{Hopkins15} code.
FIRE-2 simulations reach parsec-scale resolution to explicitly model the multiphase interstellar medium while implementing direct models for stellar evolution and feedback mechanisms.
These include energy and momentum injection from stellar winds, core-collapse and Type Ia supernovae, momentum flux from radiation pressure, photoionization and photoelectric heating.
The so-called `core' FIRE-2 suite \citep{Wetzel+23} follows the evolution of $23$ galaxies with stellar mass $M_\star$ between $10^4$ and $10^{11}\msun$ down to $z\!=\!0$, with a particular focus on MW-like systems, which form most (14/23) of the sample.
In contrast with TNG50, none of the FIRE-2 simulations model feedback from AGN, which may cause MW-like galaxies to form overly massive bulges and play a role in their elevated star formation rates at low $z$, as we discuss below.
Given their higher resolution and lack of pre-computed \hi\ fractions (see Section\,\ref{ss:building_mocks}), FIRE-2 snapshots are much more computationally demanding to process with our routines than TNG50 ones. 
For this reason, we have limited our analysis to $5$ MW-like systems from the core suite, randomly selected among those with the poorest resolution (gas particle mass of $7100\msun$), which is still more than a factor of $10$ better than TNG50.
The main properties of these systems are listed in Table \ref{t:sample_sims}.

In both TNG50 and FIRE-2, the fraction of neutral\footnote{As opposed to ionised. We clarify this as, in some different contexts, neutral and atomic hydrogen (\hi) are used as synonyms.} hydrogen associated with each gas particle ($f_{\rm n}$) is regulated by photoionization and heating from a redshift-dependent, spatially uniform ultraviolet (UV) background \citep[from][]{FaucherGiguere+09}. 
Self-shielding for high-density gas follows the prescription of \citet{Rahmati+13} in TNG50, or is directly computed using the local (averaged over the Jeans length) gas column density and opacity in FIRE-2 \citep{FaucherGiguere+15}.
Although these calculations do not account for local stellar radiation leaking from star forming regions in the host disc, we expect this to affect \hi\ fraction estimates only for column densities $N_{\rm HI}\geq10^{21}\cmmq$ \citep{Rahmati+13b, Gebek+23}.

Finally, neutral hydrogen masses must be partitioned into atomic (\hi) and molecular (H$_2$) components.
TNG50 snapshots from the MW/M31 sample have pre-computed H$_2$ fractions from three different recipes: \citet[][hereafter BR06]{BlitzRosolowsky06}, \citet[][GK11]{GnedinKravtsov11} or \citet[][K13]{Krumholz13}.
These recipes provide different approaches to the modelling of the atomic-to-molecular transition in hydrodynamical simulations. 
The BR06 recipe is based on the observed correlation between the H$_2$ fraction and the midplane pressure, whereas the other two are based on different observables, such as the gas surface density and metallicity, and are either fully analytical (K13) or calibrated on high-resolution simulations of isolated disk galaxies (GK11).
\citet{Diemer+18} provide further details on the modelling of the \hi-to-H$_2$ transition in hydrodynamical simulations.

We have applied these prescriptions also to FIRE-2 snapshots.
For the GK11 and K13 recipes, we used the \textsc{python} implementation provided by Adam R.\,H. Stevens\footnote{\url{https://github.com/arhstevens/Dirty-AstroPy}. The relevant function is \texttt{HI\_H2\_masses} using \texttt{method=2} for GK11 and \texttt{method=4} for K13}, which are detailed in \citet{Stevens+19}.
Similarly to \citet{Marinacci+17}, we have implemented the BR06 scheme by computing $R_{\rm mol}$, the ratio between the H$_2$ mass over the \hi\ mass of each gas particle/cell, as:
\begin{equation}\label{eq:BR06}
    R_{\rm mol} = (f_{\rm n}P/P_0)^\alpha
\end{equation}
where $f_{\rm n}$ is the neutral gas fraction, $P$ is the pressure associated with the gas particle/cell, $P_0=1.7\times10^4\K\cmmc$ and $\alpha=0.8$ as found by \citet{Leroy+08}.
Although the details of the implementation of the BR06 scheme vary from one study to another \citep[e.g.][]{Bahe+16,Diemer+18}, they have a very small impact on the distribution and kinematics of the low-N$_{\rm HI}$ gas that resides at the periphery of galaxy discs, where the H$_2$ fraction is typically negligible.

Fig.\,\ref{f:MH2_over_MHI} shows the ratio between H$_2$ and \hi\ total masses computed in a sphere with $100\kpc$ of radius from the centre of all simulated galaxies, and compares it with measurements from the GALEX Arecibo SDSS Survey \citep[xGASS,][]{Catinella+18}. 
The prescription of BR06 is the one that better reproduces the observed molecular-to-atomic hydrogen mass ratio, and it will be the fiducial prescription adopted for the rest of this work.
A similar conclusion was already reached by \citet{Bahe+16} studying \hi\ discs in the EAGLE simulations \citep{Schaye+15}. 
Although more recent works have preferred to use the other prescriptions \citep[e.g.][]{Ramesh+23a,Arabsalmani+23}, Fig.\,\ref{f:MH2_over_MHI} shows that these overestimate the H$_2$-to-\hi\ mass fractions by about an order of magnitude, thus we advise against their use to infer the atomic and molecular gas budget in simulated galaxies from the TNG50 and FIRE-2 suites, at least in the mass range explored here.
A $100\kpc$ aperture is adequate to encompass the very extended \hi\ disc of some of the TNG50 galaxies (e.g. 520885), but using a smaller aperture radius of $50$ or $25\kpc$ would only strengthen our results, given that the H$_2$ is more centrally concentrated than the \hi, causing $M_{\rm H2}/M_{\rm HI}$ to increase for decreasing aperture radii.
We stress that, as expected, the various recipes affect the \hi\ fractions at $N_{\rm HI}>10^{20}\cmmq$ but have no impact on the low-density CGM gas, where all neutral hydrogen is in the atomic phase (Figure \ref{f:mom0mom2_1D_MH2GK}).
In addition, we have verified the validity of a simpler partition scheme for FIRE-2 systems based on straight temperature and density cuts ($f_{\rm H2}\!=\!1$ for gas with $T\!<\!30\K$ and $n_{\rm H}\!>\!10\cmmc$, $f_{\rm H2}\!=\!0$ otherwise).
This scheme has been adopted in previous studies of FIRE-2 galaxies \citep{Orr+18,Sands+24} and leads to \hi\ masses that are larger than those determined using the \citet{BlitzRosolowsky06} recipe by $0.16$ dex.

The main properties of the simulated galaxies from Table\,\ref{t:sample_sims} are compared with those of various observational datasets in Fig.\,\ref{f:obs_vs_sim_global}.
We caution that the observational measurements in Fig.\,\ref{f:obs_vs_sim_global} are taken from the literature and make use of a variety of methods applied to different datasets.
Masses and SFRs of simulated galaxies are derived within $100\kpc$ from their centre, using the properties of the stellar particles in the snapshots.
In particular, SFRs are determined from the total birth mass of the stars younger than $100\Myr$, to match the typical timescales of the SFR tracers (UV and mid-IR) used for the MHONGOOSE galaxies \citepalias[see][]{deBlok+24}.
All FIRE-2 galaxies are located above the main sequence of star formation, as derived by \citet{Chang+15} from Sloan Digital Sky Survey \citep[SDSS][]{York+00} data, which is likely a consequence of the lack of AGN feedback in the FIRE-2 models.
TNG50 galaxies, instead, are more compatible with the SDSS, the MHONGOOSE and HALOGAS samples.
The right panel of Fig.\,\ref{f:obs_vs_sim_global} shows that the simulated galaxies have \hi\ masses compatible with those of our MHONGOOSE systems, with the $M_{\rm HI}\!-\!M_\star$ scaling relation derived by \citet{Maddox+15} combining SDSS and ALFALFA \citep{Giovanelli+05} data, and with the $M_{\rm HI}\!-\!M_\star$ relation derived by \citet{Parkash+18} from combined HIPASS \citep{Meyer+04} and WISE \citep{Wright+10} data.
In particular, the FIRE-2 galaxy m12w is characterised by a low \hi\ mass and a high SFR, very similar to J0342-47 (light-blue hexagon in Fig.\,\ref{f:obs_vs_sim_global}).
Interestingly, the $M_{\rm HI}\!-\!M_\star$ distribution of the SPARC sample of \citet{Lelli+16} appears to be systematically skewed towards lower $M_{\rm HI}$, possibly due to the instrumental differences and methods used in the two datasets.
Overall, within the limited $M_\star$ range spanned by the systems studied here ($\sim2-10\times10^{10}\msun$), and considering the broad scatter in the observational data points, we find a good agreement between simulations and observations in terms of global SFR and \hi\ masses.

\subsection{Building synthetic \hi\ observations}\label{ss:building_mocks}
The realisation of synthetic \hi\ datacubes from hydrodynamical simulations of galaxy formation has been pursued in a number of previous studies \citep[e.g.][]{Marasco+15, Oman+19}.
The procedure to build \hi\ mocks is conceptually simple, but details may vary depending on the type of simulation adopted and on the specific goals of the mocks, which is why we describe it in some detail in this Section.

Here, mock \hi\ datacubes are produced via a custom \textsc{python} software package and mimic the properties (FOV, spatial and spectral resolution, rms noise) of the MeerKAT observations.
We have preferred to build our own routines rather than adopting already existing software \citep[e.g. MARTINI,][]{Oman19,Oman24} in order to optimise the memory use and speed up the datacube construction on regular laptops.
This approach was mandatory to process the computationally demanding FIRE-2 snapshots, and was adopted for TNG50 galaxies as well for consistency.
For each simulated galaxy we produce two \hi\ cubes, corresponding to the two different resolution sets discussed in Section\,\ref{ss:mhongoose} and whose main parameters are listed in Table \ref{t:obs_setup}.
We consider all simulated galaxies as seen at a distance of $20\Mpc$, similar to the typical distance of the five MeerKAT systems studied here (see Table \ref{t:sample_obs}).
The volumes encompassed by the simulation snapshots are perfectly adequate to sample the entire FOV: TNG50 snapshots have a cubic shape with side-length of $\sim800\kpc$, FIRE-2 snapshots have a more irregular shape (because of the zoom-in nature of the suite) with a typical size of $\gtrsim1\Mpc$ and have zero contamination from low-resolution dark matter out to at least the virial radius ($\simeq200\kpc$).

In order to build the mock \hi\ datacubes we need to collect all gas particles within a snapshot that have non-zero \hi\ mass and are within the cube boundaries defined in Table \ref{t:sample_obs}.
The verification of the latter condition requires the knowledge of the galaxy centre, systemic velocity and projection on the plane of the sky.
This information is easily accessible for the TNG50 cutouts from the MW/M31 sample, which have pre-computed spatial and velocity coordinates in a reference frame where the main galaxy is at the centre, and such that the $z$-direction is parallel to the stellar angular momentum vector (that is, the galaxy is seen as face-on in the $xy$ plane).
FIRE-2 snapshots do not provide this information. 
For these simulated data we first determine the $M_\star$-weighted centroid using a series of spheres of progressively decreasing radius (starting from an initial radius of $1\Mpc$), and then use all stars within $1\kpc$ from this position to determine the galaxy centre and systemic velocity.
These are used for the computation of the stellar angular momentum vector (limited to the innermost $15\kpc$ region), which in turns sets the galactic plane.
The galaxy projection in the sky plane is regulated by two quantities: the disc inclination and position angle.
For simplicity, in this study we have adopted an inclination of $40\de$, similar to the mean inclination of the observed sample, and a position angle of $90\de$ for all simulated galaxies.
Matching the typical inclination of the observed galaxy population is particularly relevant for the study of the gas velocity dispersion, which can be affected by projection effects \citep[e.g.][]{Grisdale+17}.
Visual inspection of the stellar and \hi\ maps confirms that our projection routine works very well, especially when the stellar velocity field is clean and there are no close companions.
The desired projection is achieved by applying appropriate rotation matrixes to all gas particles within the snapshot, before the application of the spatial and velocity cuts to match the FOV and velocity range of the observations.  

The \hi\ masses of the selected gas particles are distributed within the cube according to their projected coordinates (RA, DEC, $v_{\rm LOS}$). 
However, the \hi\ mass of a given particle is not deposited entirely in a unique cube element, but it is smoothly distributed in a region of the position-velocity space.
In the spatial domain, the size of such region depends on the intrinsic resolution of the simulation, while in the velocity domain it depends on the gas thermal broadening (see below for the details).
If $M_{ijk}$ is the total \hi\ mass contained within a given cube element with indexes ($i$, $j$, $k$), then the resulting flux density $F_{ijk}$ will be given by
\begin{equation}\label{eq:M_to_I}
\frac{F_{ijk}}{\mu{\rm Jy\,beam}^{-1}} = 4.838\left(\frac{M_{ijk}}{\msun}\right) \left(\frac{\Delta v}{\kms}\right)^{-1} \left(\frac{D}{\Mpc}\right)^{-2} \left(\frac{B_{\rm maj} B_{\rm min}}{\Delta x\,\Delta y}\right),
\end{equation}
where $\Delta v$ is the channel separation, $D$ is the galaxy distance (assumed to be $20\Mpc$), $B_{\rm maj}$ and $B_{\rm min}$ are sizes of the major and minor axes of the beam, and $\Delta x$ and $\Delta y$ are the pixel sizes. 

In principle, the spatial resolution in the simulated data varies cell by cell, and can be taken as equal to $\epsilon_{\rm gas}$, the adaptive gravitational softening of a given gas cell (similar to the hydrodynamic smoothing kernel). 
However, employing a separate spatial smoothing for each gas particle in the cube is not only computationally challenging (as, for instance, up to $5\times10^6$ particles must be processed for some FIRE-2 runs), but is also very inefficient: the FIRE-2 (TNG50) simulations studied here have minimum $\epsilon_{\rm gas}$ of $1\pc$ ($72\pc$) whereas, for the adopted MeerKAT configurations, the spatial resolution in the mock data ($\epsilon_{\rm obs}$) is $\sim2\kpc$ for the HR cube and $\sim6\kpc$ for the LR one.
As a significant fraction of gas particles will have $\epsilon_{\rm gas}$ well below the resolution of the mock, we proceed as follows.
All particles with $\epsilon_{\rm gas}\!<\!\epsilon_{\rm obs}$ are smoothed using the MeerKAT beam as a kernel, modelled as a 2D Gaussian distribution. 
The particles with $\epsilon_{\rm gas}\!>\!\epsilon_{\rm obs}$ are partitioned into $N\!=\!7$ different resolution bins, logarithmically spaced in size between $\epsilon_{\rm obs}$ and the largest $\epsilon_{\rm gas}$ available. 
Particles that belong to a given bin are smoothed using a constant Gaussian kernel with FWHM set to the median $\epsilon_{\rm obs}$ of that bin.
This `adaptive' smoothing procedure is significantly faster than the ideal particle-by-particle treatment and produces very similar results.
We have experimented with varying the number of resolution bins, finding no visible differences in the final (noisy) datacube for $N>5$.

Similarly to \citet{Oman+19}, thermal broadening is accounted for by assuming a constant velocity dispersion $\sigma_{\rm th}$ of $8\kms$ for each gas particle, corresponding to a constant gas temperature of $\sim8000\K$, typical of the warm \hi\ \citep{McKee+77}.
In practice, this is achieved by convolving the mock cubes in the velocity direction using a Gaussian kernel with a standard deviation of $8\kms$.
Our treatment for the thermal broadening is justified for TNG50, which does not resolve the cold atomic and molecular phases of the ISM and does not treat gas radiative cooling below a temperature of $10^4\K$.
For consistency we use the same approach also for the FIRE-2 runs, although in principle these would permit a more sophisticated treatment thanks to their superior ISM modelling and significantly lower (down to $10\K$) temperature floor. 
However, we stress that individual \hi\ line profiles in the resulting mock data are typically a factor $\sim2$ broader than the assumed $\sigma_{\rm th}$ (Section \ref{ss:line_profiles}), which implies that the exact value adopted for this quantity has little impact on our results.

The procedure described above allows us to produce noiseless \hi\ datacubes with uniform sensitivity across the entire FoV. 
In order to transform these preliminary cubes into more realistic, MeerKAT-like mocks, two additional features must be included.
The first is the effect of the primary beam, which lowers the telescope sensitivity towards the periphery of the FoV.
This is accounted for by multiplying each velocity channel by the median primary beam map\footnote{A primary beam FWHM of $1\de$ implies that the sensitivity drops to $50\%$ at a distance of $0.5\de$ from the centre of the FoV. For simplicity, we ignore the small variations of the primary beam with the wavelength.}, which is available as part of the observational pipeline.
The second feature is the injection of noise, which is done under the simplifying assumption that it is uncorrelated with the signal, in which case it is sufficient to produce a separate `noise-cube' that can be summed to the (primary-beam degraded) mock data.
Under the assumption of a uniform, Gaussian noise \citep[which holds up in MHONGOOSE data to a $4\sigma$ level, see][]{Veronese+24}, this noise-cube can be produced starting from a temporary 3D array, having the same size of the data, made by values randomly extracted from a Gaussian distribution with zero mean and an arbitrary standard deviation. 
This temporary array is then convolved with the telescope beam\footnote{Under the assumption of a Gaussian beam, cube elements resulting from this step still follow a Gaussian distribution}, re-normalised so that its resulting standard deviation is equal to that desired (that is, $0.15\miJyb$ for the HR and $0.25\miJyb$ for the LR), and finally summed to the mock data.  
We save the original noiseless cubes and the final MeerKAT-like realisations separately, for future comparison.

\section{Results}\label{s:results}
\subsection{Detailed \hi\ properties of a simulated galaxy} \label{ss:single_galaxy}
\begin{figure*}
\begin{center}
\includegraphics[width=0.95\textwidth]{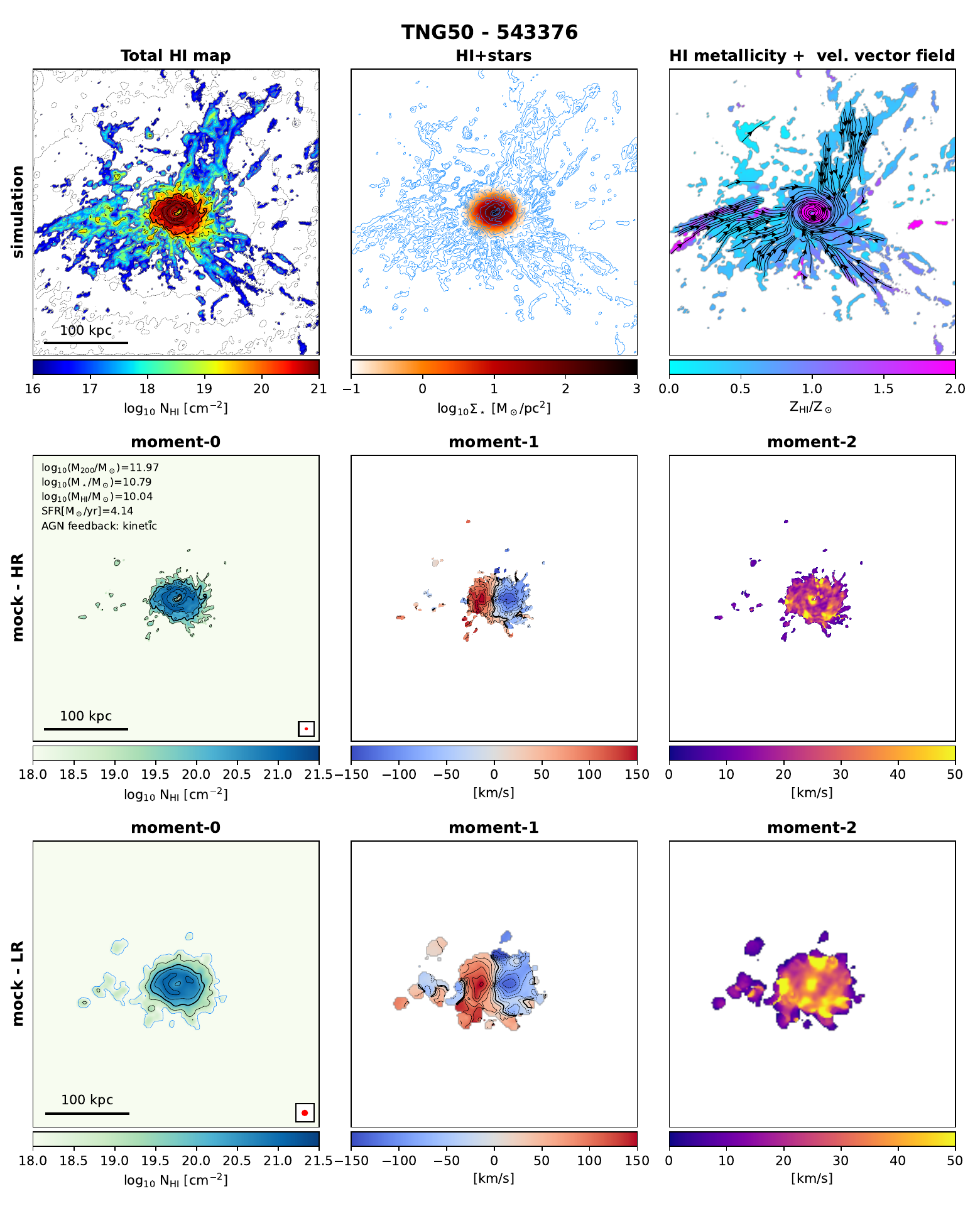}
\caption{Diagnostic maps of galaxy 543376 in the TNG50 simulation. \emph{Top row:} \hi\ column density map down to a depth of $10^{16}\cmmq$ (left), stellar surface density map overlaid with \hi\ contours (middle), \hi\ metallicity map overlaid with the vector velocity field. \emph{Central and bottom rows:} moment maps for the HR (central) and LR (bottom) mock \hi\ data, derived with SoFiA. The main galaxy properties are listed in the middle-left panel.
Iso-contours in $N_{\rm HI}$ maps are drawn at levels of $10^{16}$, $10^{17}$ (in grey, top panel alone), $10^{18}$ (in blue, top and bottom panels), $10^{19}$, $10^{20}$, $10^{20.5}$ and $10^{21}\cmmq$ (in black, all panels), with the iso-contour at $10^{20}\cmmq$ highlighted with a thicker line. In moment-1 maps, the contour at the systemic velocity is shown by a a black line, and consecutive contours are spaced by $25\kms$.
Beam sizes are shown as red ellipses in the bottom-right corner of moment-0 maps.
The dashed grey iso-contours in the top-left panel trace the surface density distribution of the dark matter halo.}
\label{f:TNG50_maps_example}
\end{center}
\end{figure*}

Before presenting a quantitative comparison between the simulated and the observed galaxies in a statistical manner, in this Section we discuss the detailed \hi\ properties of a single simulated galaxy, system 543376 in TNG50.
The goal is to provide a visualisation of the main differences between what the hydrodynamical simulation predicts and what could be determined with a putative \hi\ observation with the MeerKAT telescope. 
Also, this exercise allows us to highlight the differences between the two resolution cases considered.

Fig.\,\ref{f:TNG50_maps_example} shows a collection of diagnostic maps for 543376\footnote{Maps analogous to those of Fig.\,\ref{f:TNG50_maps_example} for the remaining simulated galaxies are stored at \href{https://drive.google.com/drive/folders/1eOslF317JGeLkry07oUgDrDNaiZoLYVv?usp=sharing}{this repository}.}.
The top panels of Fig.\,\ref{f:TNG50_maps_example} show properties that are predicted from the simulation but cannot be readily derived observationally, such as an \hi\ column density map down to $N_{\rm HI}$ of $10^{16}\cmmq$ (top-left panel) and the \hi-phase metallicity map with the vector velocity field overlaid on top (top-right panel).
These maps show that 543376 is characterised by an extended \hi\ disc with a diameter of $\sim60\kpc$ (measured at the \hi\ column density level of $10^{20}\cmmq$), surrounded by a few extended \hi\ filaments that connect the galaxy with the IGM, reaching the edge of the simulated FoV.
The filaments are made of low-density \hi\ ($N_{\rm HI}\lesssim10^{19}\cmmq$) that streams towards the galactic disc, as can be seen from the 3D velocity field.
The lack of clear stellar counterparts and the overall low metallicity indicate that the filaments are likely tracing cold gas accretion from the cosmic web, although part of the infalling gas may originate from re-accretion of material previously ejected from the disc by stellar feedback (recycling), especially the regions with higher metallicity (such as for the leftmost \hi\ stream).
It is not unusual for MW-like galaxies in cosmological hydrodynamical simulations to be embedded in multiple confluent filaments \citep[e.g.][]{Keres+09a,vandeVoort+19}, although this is not necessarily the case for all of the simulated systems considered in this work.
We stress that the detailed \hi\ morphology of this system is well resolved in the simulations. For instance, the entire top-right filament is sampled by $\sim10^4$ gas particles with \hi\ mass fraction $>10^{-5}$, and even the smallest isolated clump visible in the total \hi\ map is sampled by at least a few tens particles.

The panels in the central and bottom rows of Fig.\,\ref{f:TNG50_maps_example} show the standard moment-0, moment-1 and moment-2 maps derived by processing the mock \hi\ datacubes (which include noise) using the same methodology that was adopted for the MeerKAT galaxies, which is described in Section\,\ref{ss:mhongoose}.
As expected, the differences with respect to the `infinite-depth' N$_{\rm HI}$ map are striking, especially for the HR case (central row) where the \hi\ filaments surrounding the galaxy disc are completely invisible.
The increased sensitivity of the LR case permits instead to capture a few faint \hi\ clouds located to the left of the galactic disc and, importantly, to marginally increase the size of the \hi\ disc itself, emphasising the outer warp in the velocity field (moment-1 map), visible as a `twist' in the kinematic position angle.
A possible interpretation for kinematic \hi\ warps observed in real galaxies is that they originate from accretion events \citep[e.g.][]{JiangBinney99}, which is likely the case here as well.
We point out that the degraded resolution of the LR case prevents us to identify the filamentary \hi\ structure that would be otherwise partially visible with a sensitivity of $10^{18}\cmmq$ at a higher resolution, as shown by the top-left panel in Fig.\,\ref{f:TNG50_maps_example}.

Another relevant feature of the mock observations is the presence of localised, multiple high-velocity peaks in the moment-2 maps, visible as bright spots at velocity dispersion $\sigma\gtrsim50\kms$ for both the HR and the LR cases.
Such large values for $\sigma$ are very rarely observed in MeerKAT galaxies (see Fig.\,\ref{f:MHONGOOSE_momentmaps}), or in local star-forming galaxies \citep[e.g.][]{Tamburro+09, Bacchini+19}.
In general, high-$\sigma$ peaks can originate from kinematically hotter components (i.e., highly turbulent \hi\ regions), or can be caused by the overlap of multiple cold kinematic components along the line of sight.
These two options are further discussed in Section \ref{ss:line_profiles}, where we characterise the shape of the \hi\ line profiles in our mock datacubes and compare it to the observations.
Interestingly, for galaxy 543376, these high-$\sigma$ peaks appear to be localised in regions where the \hi\ filaments and the disc join each other (in projection), which would support a physical origin from the dissipation of gravitational energy of the accreting gas \citep[e.g.][]{KlessenHennebelle10,Krumholz+18}.
We have verified that this is the case by visually inspecting a face-on projection of all individual \hi-rich gas particles, confirming that particles at the intersection regions between the filaments and the disc show a significantly larger $|v_{\rm z}|$ (the modulus of the velocity in the direction perpendicular to the disc) with respect to the surrounding gas. This experiment confirms that projection effects do not play a major role.
Visual inspection of the diagnostic maps of the other simulated galaxies reveals that the presence of high-$\sigma$ regions is a common feature of TNG50 and FIRE-2 simulated galaxies.
In Section\,\ref{ss:mom2_feedback}, we discuss how the combined effects of feedback and accretion from the cosmic web play an important role in producing these features.

In the remaining part of the analysis we will focus on a comparison with the observed sample in terms of moment-0 and moment-2 maps. 
The study of velocity fields is extremely interesting as it can potentially reveal key information on the properties of the accreting gas and of dark matter halos \citep[e.g.][]{Marasco+18,Oman+19,Sharma+22,Roper+23}, but it requires a dedicated analysis which we plan to present in a future contribution.

\subsection{Column density and velocity dispersion distributions}\label{ss:mom0mom2}
\begin{figure*}
\begin{center}
\includegraphics[width=0.9\textwidth]{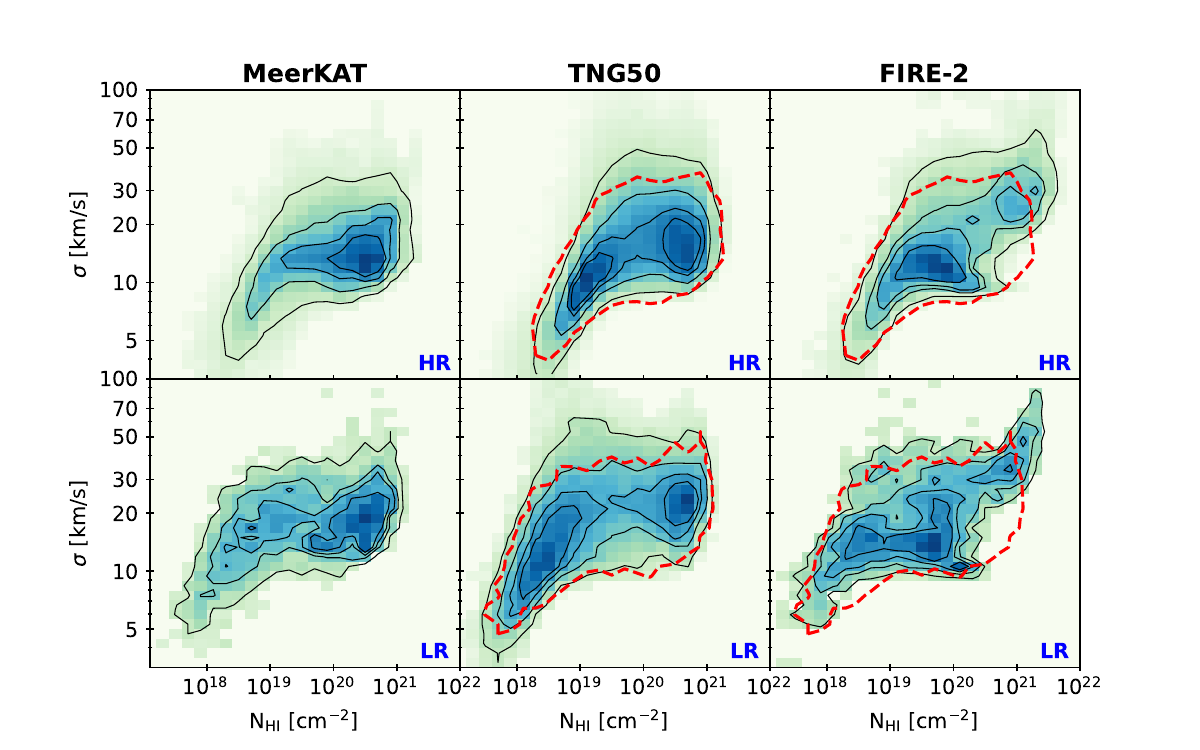}
\caption{Two-dimensional density histograms of the inclination-corrected \hi\ column density and second moment values for MeerKAT (left panels), TNG50 (central panels) and FIRE-2 (right panels) galaxies. The top and bottom panels show the HR and LR cases, respectively. The black contours encompass $25\%$, $50\%$, $75\%$ and $95\%$ of the highest density gas of the integrated distribution. The red-dashed contours in the TNG50 and FIRE-2 panels reproduce the outermost contour of the MeerKAT distribution, and are included as a reference.
}
\label{f:mom0mom2}
\end{center} 
\end{figure*}

\begin{figure*}
\begin{center}
\includegraphics[width=0.73\textwidth]{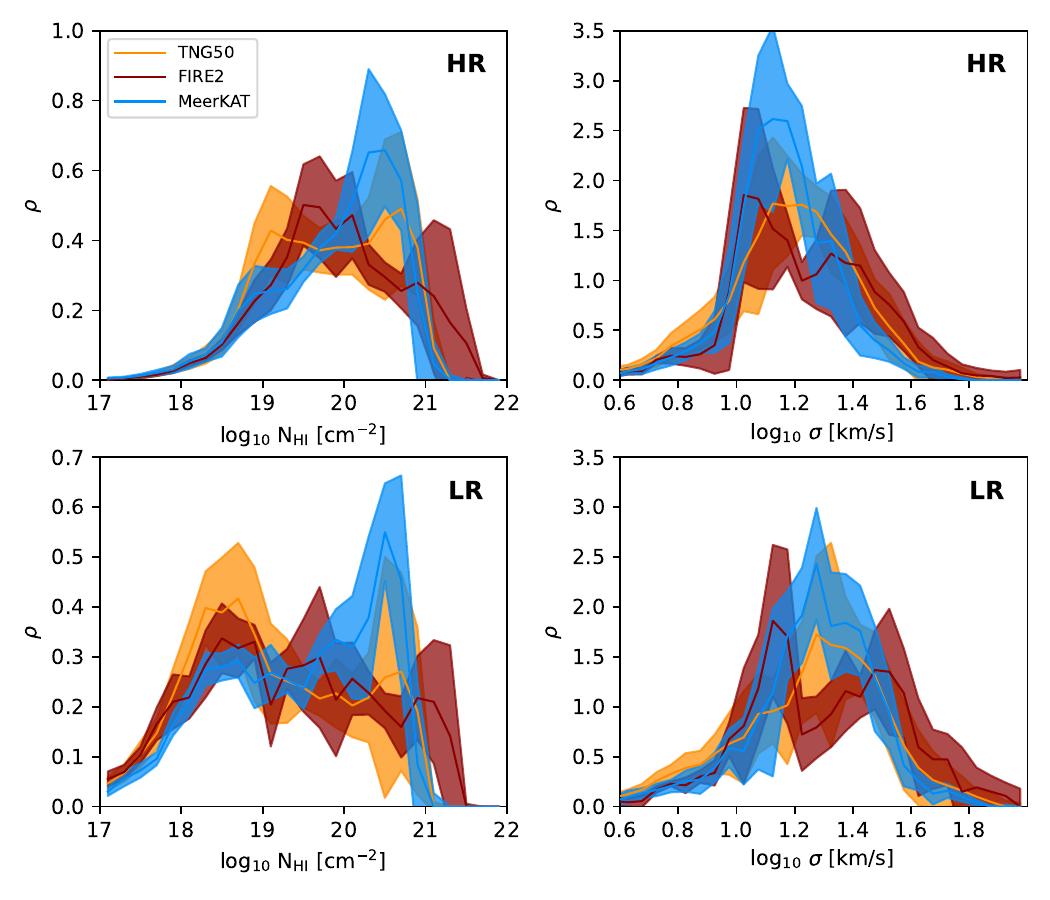}
\caption{Probability density distribution of inclination-corrected $N_{\rm HI}$ (left panels) and second (right panels) moment values for MeerKAT (blue), TNG50 (orange) and FIRE-2 (dark red curves) galaxies. The HR and LR cases are shown in the top and bot tom panels, respectively. Solid lines show the median distribution for a given galaxy sample, while shaded regions show the scatter within the sample, computed as the difference between the 84th and 16th percentiles.
}
\label{f:mom0mom2_1D}
\end{center}
\end{figure*}

We now compare the observed \hi\ properties of the three galaxy samples studied in this work. 
To do so, we rely on two-dimensional density histograms of the zeroth and second moment values, which we compute for all galaxies combined in a given sample, using only pixels within the masks provided by SoFiA. It is worth stressing again that the masks were constructed in an identical manner for all three samples.
The 2D distributions are shown in Fig.\,\ref{f:mom0mom2}, with a separate panel for each sample and resolution mode, and are produced using bins of $0.2$ dex in $N_{\rm HI}$ and $0.05$ dex in $\sigma$.
To account for projection effects, the column density inferred from moment-0 maps were corrected for galaxy inclination, which we take from Table 1 of \citetalias{deBlok+24} for the MHONGOOSE galaxies, from \citet{Stuber+23} for J0342-47, and is fixed at $40\de$ for the simulated samples.

In our observed sample, the 2D distribution resembles that presented in \citetalias{deBlok+24}, though with slightly lower S/N due to the smaller number of galaxies analysed here. 
We clearly see a well-defined peak at $N_{\rm HI}\simeq5\times10^{20}\cmmq$ and $\sigma\simeq15\kms$ for both resolution cases. 
The typical $\sigma$ stays constant down to $N_{\rm HI}$ values in the range of $10^{19}$--$10^{20}\cmmq$ (the exact value depends on the resolution), and below $10^{19}\cmmq$ the two quantities appear to be strongly correlated.
As extensively discussed in section\,7.3 of \citetalias{deBlok+24}, this correlation is artificial, and is caused by the limited sensitivity: low signal-to-noise \hi\ lines will have their high-velocity wings clipped out by the SoFiA mask, which will diminish the measured line width in a way that depends on the strength of the signal.
This also explains why $\sigma$ stays flat down to column densities that are lower in the LR mode.

Although similar in its overall shape, the 2D distribution determined for the simulated galaxies shows important differences with respect to the MeerKAT sample. 
First, both TNG50 and FIRE-2 galaxies show at least two separate peaks, one typically associated with low values of $N_{\rm HI}$ and $\sigma$, the other associated with higher densities and line-widths.
This bimodality is particularly evident in TNG50 galaxies for the LR mode, but it is present also in the other simulated samples. 
We have verified that the bimodality is present regardless of the recipe adopted to separate molecular from atomic hydrogen, thus we can consider it as a systematic feature of the hydrodynamical models analysed here. 
In MeerKAT, the bimodality is not visible in the HR cubes. 
A very weak secondary peak is visible in the LR cubes at $N_{\rm HI}\!\sim\!10^{19}\cmmq$ \citepalias[as also noticed by ][]{deBlok+24}, but this is much fainter than that in the simulated galaxy sample, and shifted to a factor $\sim10$ higher column density.
A consequence of this discrepancy is that the simulated galaxies show an excess of low-$N_{\rm HI}$ material, as we discuss further below.
A second difference that can be appreciated from the panels in Fig.\,\ref{f:mom0mom2} is that the simulated galaxies feature a more pronounced extension to high-$\sigma$ values than the observed systems.
We have already mentioned this feature when discussing the detailed \hi\ properties of 543376 in Section\,\ref{ss:single_galaxy}.
Here, we notice that the high-$\sigma$ \hi\ in the simulated galaxies is present at all column densities, but it is more visible at $10^{19}\!<\!N_{\rm HI}\!<\!10^{20}\cmmq$ in TNG50 and at $N_{\rm HI}\!>\!10^{20.5}\cmmq$ in FIRE-2.
This suggests that the high-$\sigma$ regions are more concentrated within the densest, star-forming regions of FIRE-2 galaxies, whereas they are distributed at the periphery of the discs of TNG50 galaxies, as we have seen for the case of 543376.
At a column density of $\sim10^{20}\cmmq$, we observe a bifurcation in the velocity dispersion of FIRE-2 galaxies, which can be linked to the multiple components present in these simulations, as discussed later in the paper (Section \ref{ss:line_profiles}). This bifurcation appears in the LR plot but is less visible in the HR plot.

Fig.\,\ref{f:mom0mom2_1D} shows the probability density function (pdf) of the inclination-corrected $N_{\rm HI}$ and second moment ($\sigma$) values, averaged over all galaxies in a given sample, for the three samples and the two resolutions considered. 
Specifically, for a given sample, the median pdf is shown as a solid line, while galaxy-to-galaxy fluctuations are represented as shaded regions and are computed as the difference between the 84th and 16th percentile of the ensemble of pdfs.
Fig.\,\ref{f:mom0mom2_1D} better highlights the differences between the observed and simulated systems.
The $N_{\rm HI}$ distributions of the MeerKAT galaxies and of the simulated ones have a markedly different shape, indicating that, on spatial scales of a few kpc, there is an intrinsic difference between the observations and the simulations in the distribution of \hi\ inside and around local star forming galaxies.
In particular, both the TNG50 and the FIRE-2 systems show an excess of \hi\ at column densities $\lesssim10^{20}\cmmq$, which we further investigate in Section \ref{ss:HIspatialdistrib}.
FIRE-2 systems also show an excess of \hi\ at $N_{\rm HI}>10^{21}\cmmq$.
Such high-N$_{\rm HI}$ values are rarely observed in real galaxies, and in the simulations are largely dependent on the atomic-to-molecular partition recipe adopted. 
They would disappear if we used a recipe different from that of \citet{BlitzRosolowsky06}.
This is shown in Fig.\,\ref{f:mom0mom2_1D_MH2GK} for the recipe of \citet{GnedinKravtsov11}, but the recipe of \citet{Krumholz13} gives comparable results.
We remark though that the \citet{BlitzRosolowsky06} recipe is the one that better reproduces the observed molecular gas fraction in galaxies, at least in the mass range considered in this study, as shown in Fig.\,\ref{f:MH2_over_MHI}.
The right panel of Fig.\,\ref{f:mom0mom2_1D} shows that the differences in the $\sigma$ distribution are significant.
First, with respect to the observations, the simulations show a more pronounced tail towards large $\sigma$ values.
To provide a reference, it is $2$ ($4$) times more likely to find a pixel with $\sigma$ of $\sim40\kms$ in the moment-2 maps of TNG50 (FIRE-2) galaxies than in the real ones.
Additionally, while the $\sigma$ distributions in the MeerKAT sample and the TNG50 sample are unimodal and peak at approximately the same value ($\sim15\kms$), FIRE-2 galaxies feature a bimodal distribution with separate peaks at $\sim12\kms$ and $\sim30\kms$.
As shown in Fig.\,\ref{f:mom0mom2}, these two $\sigma$ peaks correspond to different $N_{\rm HI}$ regimes, with the high-density and low-density \hi\ showing distinct kinematics.
The comparison presented here highlights how the differences between the observed and simulated galaxy samples are not negligible, especially in the case of FIRE-2.

\subsection{\hi\ spatial distribution}\label{ss:HIspatialdistrib}
\begin{figure}
\begin{center}
\includegraphics[width=0.4\textwidth, trim={0.0cm 0.0cm 0.0cm 0.8cm},clip]{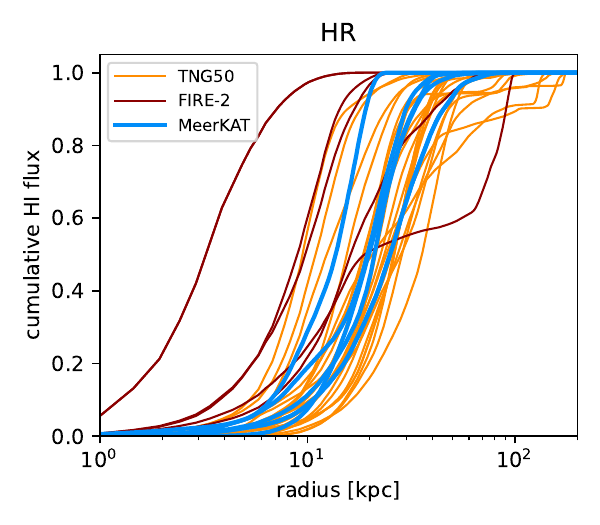}
\caption{Cumulative \hi\ flux as a function of the projected distance from the galaxy centre for individual MeerKAT (blue), TNG50 (orange) and FIRE-2 (dark red) systems (HR case). Abrupt discontinuities due to the presence of isolated \hi-rich satellites have been removed from all profiles.}
\label{f:cumulative_HR}
\end{center}
\end{figure}

\begin{figure}
\begin{center}
\includegraphics[width=0.4\textwidth, trim={0.0cm 0.0cm 0.0cm 0.8cm},clip]{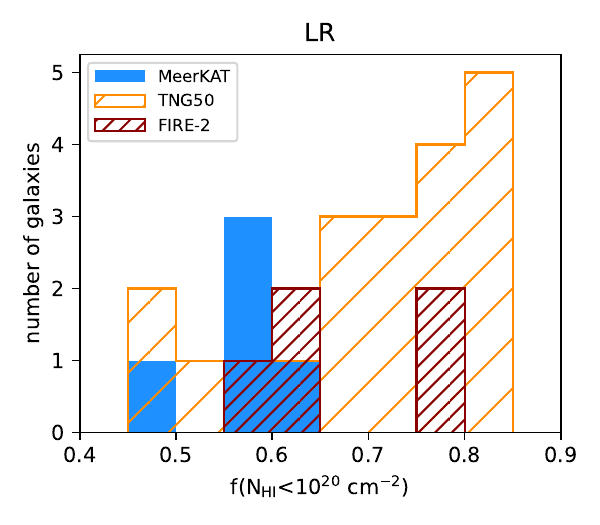}
\caption{Distribution of the fraction of spaxels with \hi\ column densities below $10^{20}\cmmq$ for the LR cubes in MeerKAT (filled blue histogram), TNG50 (hatched orange histogram) and FIRE-2 (hatched dark red histogram). The fraction is computed with respect to all spaxels that contribute to the moment maps.}
\label{f:fHI_faint_LR}
\end{center}
\end{figure}

\begin{figure*}
\begin{center}
\includegraphics[width=\textwidth,trim={0 4.0cm 4.0cm 0},clip]{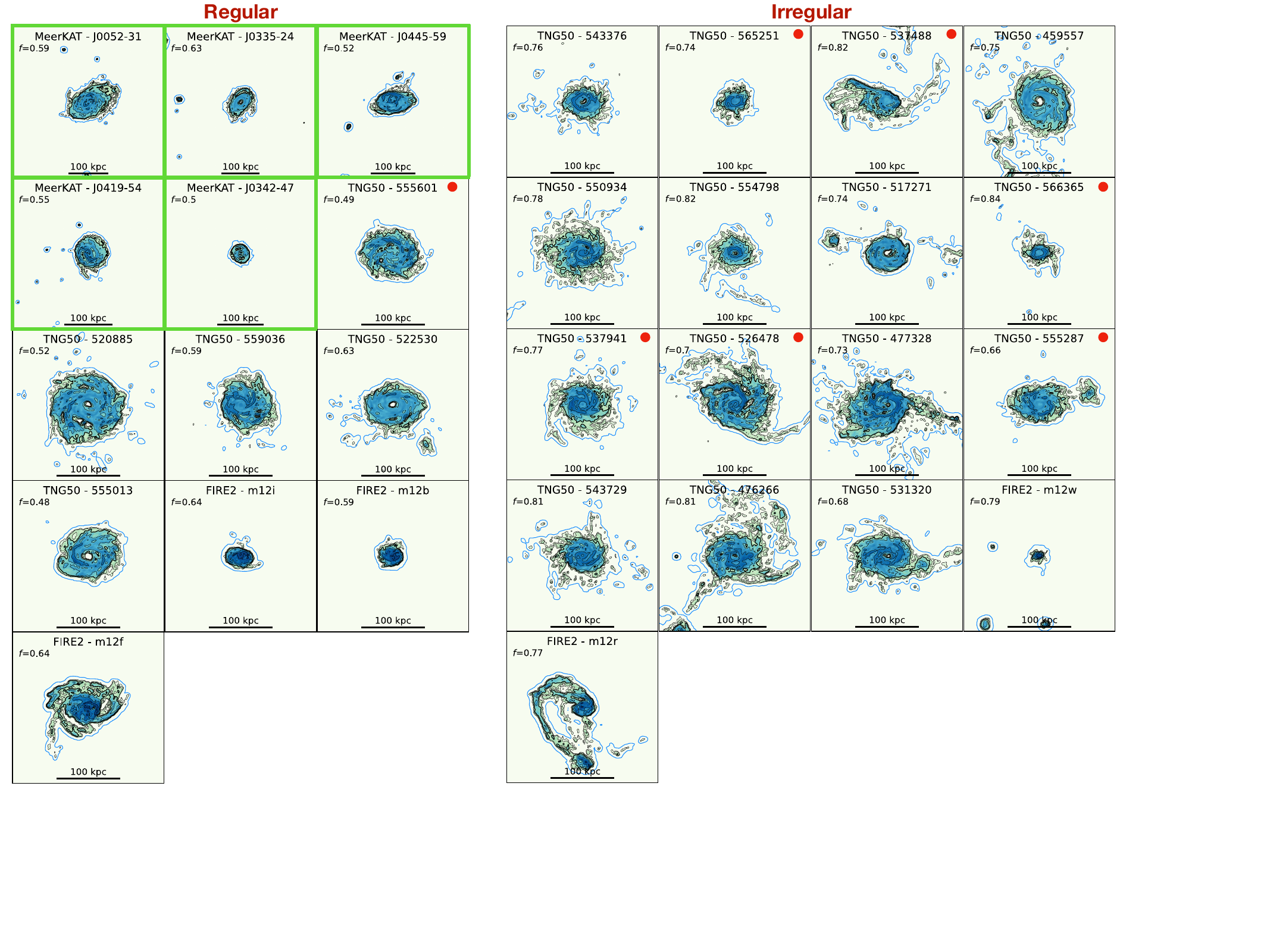}
\caption{Moment-0 maps for all galaxies in our sample divided into regular (on the left) and irregular (on the right), depending on whether the fraction $f$ of spaxels with \hi\ column density below $10^{20}\cmmq$ is higher or lower than $0.65$ (see Fig.\,\ref{f:fHI_faint_LR}). $f$ values of individual galaxies are annotated in each panel. Red dots mark TNG50 galaxies that have AGN feedback in thermal mode. Black iso-contours from the HR maps are drawn at levels of $10^{19}$, $10^{20}$ (thicker line), $10^{20.5}$ and $10^{21}\cmmq$. An additional iso-contour from the LR maps at $10^{18}\cmmq$ is shown in blue.
Green boxes are used to highlight galaxies observed with MeerKAT. The maps were derived with SoFia-2, using the same parameters for all galaxies.}
\label{f:regular_vs_diffuse_mom0}
\end{center}
\end{figure*}

\begin{figure}
\begin{center}
\includegraphics[width=0.5\textwidth]{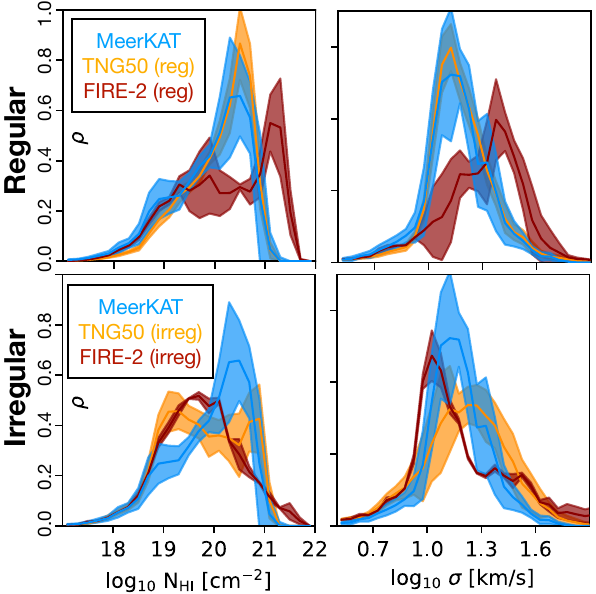}
\caption{As in Fig.\,\ref{f:mom0mom2_1D} for the HR case, but with the simulated galaxies split into regulars (top panels) and irregulars (bottom panels). Regular TNG50 galaxies follow the distributions observed with MeerKAT very well.}
\label{f:regular_vs_diffuse_NHI_sigma}
\end{center}
\end{figure}

While Section\,\ref{ss:mom0mom2} focused on the overall column density and velocity dispersion distribution for the ensemble of galaxies in a given sample, we now address the question of whether individual TNG50 and FIRE-2 systems show any relevant differences with respect to the observed galaxies in terms of spatial distribution of their \hi\ content.

Fig.\,\ref{f:cumulative_HR} shows the cumulative \hi\ profiles (normalised to $1$) as a function of the projected distance from the galactic centre for each galaxy in our samples.
We show the profiles for the HR case, but the LR case is very similar.
To avoid abrupt discontinuities caused by the presence of isolated \hi-rich satellites, we have assumed that a profile has reached its asymptotic value if there exists a window with a size of at least $20\kpc$ where it shows no growth.
In practice, each profile is inspected for the presence of a flat region with a length of at least $20\kpc$ and, if such a region is identified, the remaining part of the profile at larger radii is discarded. 
This approach allows us to trace the distribution of \hi\ that is spatially connected with the main galaxy, ideally avoiding to sample external galaxies that may be dynamically unrelated with the observed target.

Fig.\,\ref{f:cumulative_HR} indicates that FIRE-2 galaxies are typically too concentrated compared to the TNG50 and the observed systems.
On the contrary, several TNG50 systems are less concentrated than the observed galaxies as they often show central \hi\ holes (visible in Fig.\,\ref{f:regular_vs_diffuse_mom0}) related to the ejective character of SMBH feedback, as discussed in \citet{Gebek+23}.
While we refer the reader to \citet{Diemer+19} and \citet{Gensior+24} for a detailed investigation of the \hi\ mass-size relation in MW-like galaxies from hydrodynamical simulations, it certainly appears that, at least for the samples considered in our study, there is a large discrepancy between FIRE-2 and TNG50 in the predicted sizes of \hi\ discs at fixed \hi\ mass.
In addition, most TNG50 systems have a significant fraction of \hi\ distributed at large distances ($50\!-\!150\kpc$) from the galactic centre.
In particular, the percentage of \hi\ flux located beyond a projected distance of $70\kpc$ from the galaxy centre is always $<1\%$ for all systems in the MHONGOOSE sample, while it can reach up to $14\%$ in the TNG50 sample and $32\%$ in one FIRE-2 galaxy (m12r).
This result does not depend strongly on the length of the window used to filter out \hi-rich satellites, and indicates a general mismatch in the way the cold gas is spatially distributed in the simulated galaxies and in the observed ones.

The fact that several systems feature too much \hi\ in their outskirts is not due to the presence of a collection of high-density clumps at large galactocentric distances, but rather to an excess of low-$N_{\rm HI}$ material surrounding the galaxy discs.
This can be appreciated by looking at the diagnostic maps in Appendix \ref{a:supplementary}, and is quantified in Fig.\,\ref{f:fHI_faint_LR}, which shows how the fraction of spaxels at $N_{\rm HI}\!<\!10^{20}\cmmq$ (computed with respect to all spaxels available within the SoFiA mask, without any filtering) is distributed amongst the different galaxy samples, for the LR case which is the most sensitive to low-$N_{\rm HI}$ features.
The percentage of low-$N_{\rm HI}$ spaxels is in the range $45\!-\!65\%$ for all MeerKAT galaxies. Most galaxies in the simulated sample have values higher than the highest MeerKAT value: two FIRE-2 and 15 TNG50 galaxies show values between $65\%$ and $85\%$.
A similar result can be derived for the HR mode, but with a marginally less clean separation between the observed and simulated samples due to the lower N$_{\rm HI}$ sensitivity, which prevents the identification of the faintest \hi\ features in the outskirts of the simulated galaxies.
We stress that, as shown by \citet{Veronese+24}, the stacking of MHONGOOSE data suggests the absence of additional \hi\ signal in the galaxy halos down to column densities of $\sim10^{17}\cmmq$.

\subsection{Regular and irregular \hi\ discs}
\label{ss:regular_vs_irregular}
The separation visible in the histograms of Fig.\,\ref{f:fHI_faint_LR} allows us to split galaxies in our samples into two categories, above and below $f({\rm N}_{\rm HI}\!<\!10^{20}\cmmq)$ values of $0.65$.
The total \hi\ maps of galaxies belonging to each category are collected in Fig.\,\ref{f:regular_vs_diffuse_mom0} for the HR case, and appear quite different.
Galaxies with $f<0.65$ generally show a regular morphology characterised by well-defined discs, possibly surrounded by a small number of \hi-rich satellites. 
All MeerKAT galaxies (highlighted with green boxes in Fig.\,\ref{f:regular_vs_diffuse_mom0}) fall in this category.
Galaxies with larger $f$ values show more irregular morphologies, featuring filamentary or uneven \hi\ envelopes which suggest a more complex interplay with their environment.
Moment-1 and moment-2 maps for the two categories are shown in Appendix \ref{a:supplementary}.
A visual inspection of these maps indicates that, qualitatively, the simulated galaxies that belong to the regular category (especially those from the TNG50 sample) are very similar to the observed galaxies in their morphology and kinematics.
We explore this more quantitatively in Fig.\,\ref{f:regular_vs_diffuse_NHI_sigma}, where we present again the N$_{\rm HI}$ and $\sigma$ distributions shown in Fig.\,\ref{f:mom0mom2_1D} for the HR case, but for the regular and irregular simulated systems separately.
Interestingly, in TNG50, although regular galaxies make up to only one fourth of the simulated sample, their typical N$_{\rm HI}$ and $\sigma$ distributions are in excellent agreement with the observations.
This is not the case for the irregular TNG50 galaxies, or for FIRE-2 galaxies in general.
Insights on the physical origin of the difference between the regular and irregular populations are provided below, and in particular in Section \ref{ss:regular_vs_irregular_origin}.

\section{Discussion}\label{s:discussion}
In this Section we focus preferentially on the simulated galaxies and discuss how some of their properties (environment, star formation rates, gas accretion rates) play a role in shaping their \hi\ distribution and kinematics.
We also attempt to quantify the complexity of their \hi\ profiles in more detail, in order to provide more insightful comparison with the data.

\subsection{Gas inflows and outflows}\label{ss:inflow_outflow}
\begin{figure*}
\begin{center}
\includegraphics[width=0.9\textwidth]{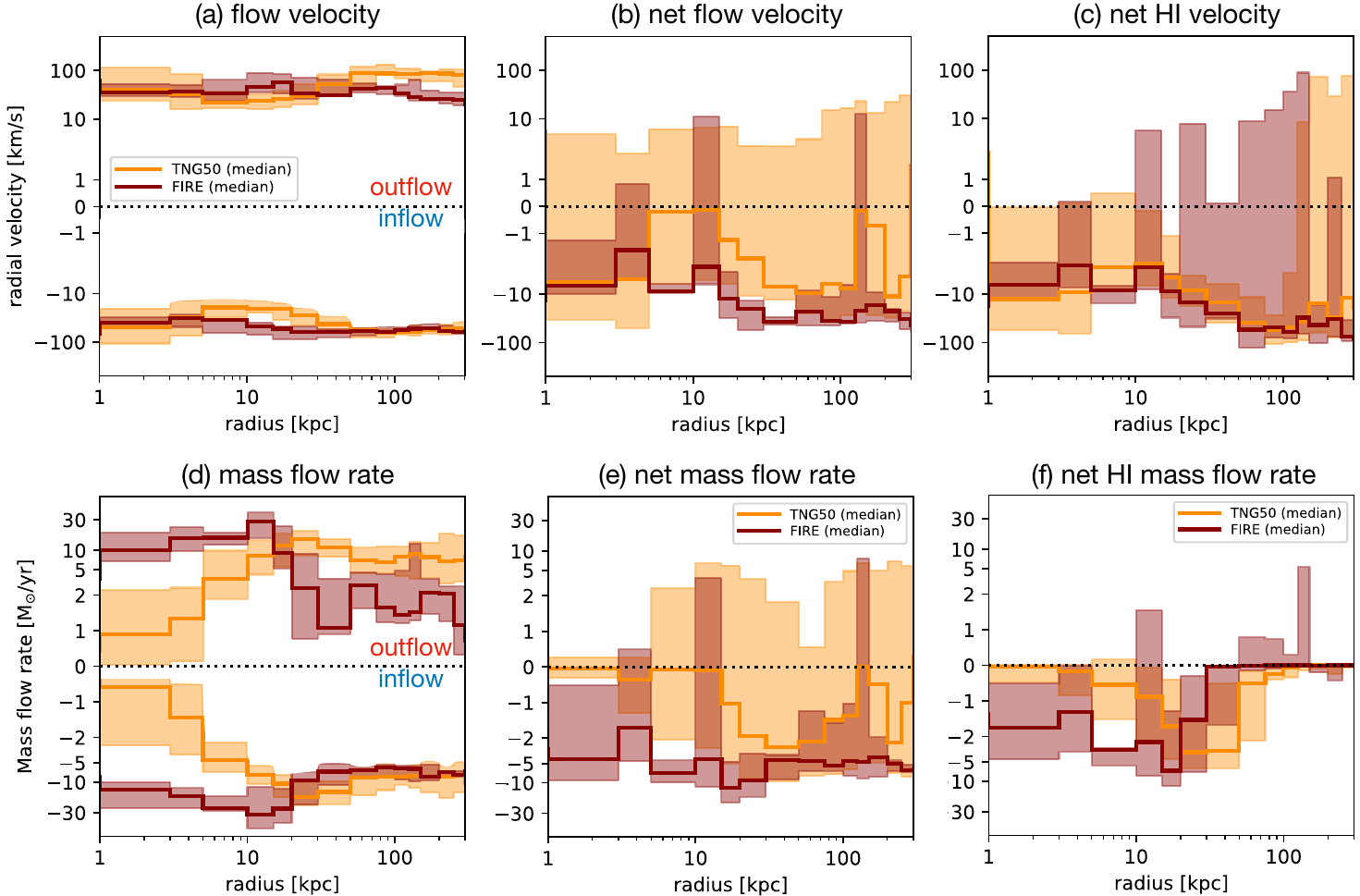}
\caption{Radial velocities (top panels) and gas mass flow rates (bottom panels) computed in spherical concentric shells, as a function of the shell radius, for the TNG50 (orange) and FIRE-2 (dark red) galaxies.
Panels (\emph{a}) and (\emph{d}) show the inflow and outflow components separately, panels (\emph{b}) and (\emph{e}) show the net (outflow-inflow) flow for all gas phases, panels (\emph{c}) and (\emph{f}) show the net flow for the \hi\ alone. Thick lines show the median trend, while the shaded area show the difference between the 84th and 16th percentile. Negative values indicate inflow.}
\label{f:gas_flows}
\end{center}
\end{figure*}

\begin{figure}
\begin{center}
\includegraphics[width=0.35\textwidth]{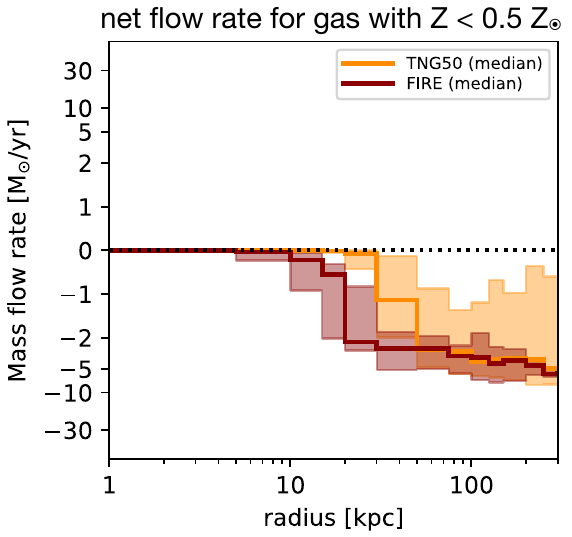}
\caption{As in panel (\emph{e}) of Fig.\,\ref{f:gas_flows}, but for all gas with $Z<0.5\zsun$.}
\label{f:gas_lowZ_flows}
\end{center}
\end{figure}

One of the great advantages of hydrodynamical simulations of galaxy formation and evolution is that the three-dimensional gas flow can be traced at any distance from the galaxy centre, regardless of the gas temperature, density and ionisation state.
Accurate measurements for the gas inflow and outflow rates are very difficult to obtain observationally \citep[e.g.][]{Sancisi+08, DiTeodoroPeek21, Kamphuis+22, Marasco+23a} due to a combination of projection effects and the impossibility of sampling all gas phases simultaneously in a homogeneous way, with sufficiently high spatial and spectral resolution.

To investigate the instantaneous gas flow in the simulated galaxies we have divided each snapshot into a series of spherical shells, all centred at the galactic centre and with variable thickness. 
In each shell, we have computed the mass-weighted radial speed $\avg{v_{\rm r}}$ and the net flow rate $\dot{M}$, which are defined as
\begin{align}
\avg{v_{\rm r}} & = \frac{\sum_{n=1}^{N} (v_{{\rm r},\,n}\,m_n\,w_n)}{\sum_{n=1}^{N}  (m_n\,w_n)}, \label{eq:radial_speed}
\\
\dot{M} & =\frac{1}{\Delta r} \sum_{n=1}^{N}  (v_{{\rm r},\,n}\,m_n\,w_n), \label{eq:massflowrate}
\end{align}
where the sum is extended to all $N$ gas particles in the shell, $v_{{\rm r},\,n}$ is the particle radial speed, $m_n$ is the particle mass, $w_n$ is a weight that depends on which gas phase is being considered (it is equal to the gas particle \hi\ fraction if we focus on the \hi, or to $1$ if we consider all phases together), and $\Delta r$ is the shell thickness.
For TNG50 we have also included the so-called `wind particles' \citep[see][]{Pillepich+18} assuming that they are made of fully ionised gas, finding that their contribution to the flow rate is negligible for the galaxies studied here.
Equation\,(\ref{eq:radial_speed}) gives the typical speed at which the gas is flowing into (or out from) the shell considered, while Equation\,(\ref{eq:massflowrate}) provides the instantaneous mass flow rate in a shell.
Both equations can be applied to the inflowing ($v_{{\rm r},\,n}\!<\!0$) and outflowing ($v_{{\rm r},\,n}\!>\!0$) gas particles separately in order to study the inward and outward motions independently, or to all particles (regardless of the sign of their $v_{{\rm r},\,n}$) to infer the properties of the `net' gas flow. By averaging the inwards and outwards flows, the net flow indicates whether the gas in a given shell is moving on average in one direction or the other.
We remark that, as in our calculations we account for all gas particles available in the snapshot, at small radii ($r<10-20\kpc$) the flow will be dominated by motions within the galaxy disc.

Fig.\,\ref{f:gas_flows} provides the radial profiles of $\avg{v_{\rm r}}$ (top panels) and $\dot{M}$ (bottom panels) for the inflowing and outflowing components separately (leftmost panels), for the net flow of all gas phases (central panels) and for the net flow of the \hi\ phase alone (rightmost panels).
Individual galaxies show a substantial scatter in their radial profiles; thus we focus on the median profiles computed (shell by shell) using all galaxies in the TNG50 and FIRE-2 samples separately, which are represented by the thick lines in Fig.\,\ref{f:gas_flows}.
The leftmost panels of Fig.\,\ref{f:gas_flows} show that, at all radii, the gas moves inwards and outwards with velocities ranging from a few tens up to $\sim100\kms$ (panel \emph{a}), leading to flow rates up to a few tens $\msunyr$ in both directions (panel \emph{d}).
Our interpretation for the remarkable symmetry shown by the inflow and outflow rate profiles is that gas in the halo moves inwards following slightly eccentric orbits, possibly caused by a triaxial gravitational potential, thus measurements via Equation (\ref{eq:massflowrate}) that assumes a spherical symmetry will infer large inwards and outwards rates due to the gas radially oscillating along its non-circular orbits.
We have verified that the percentage of gas mass that flows in the tangential direction (as opposite to the gas that flows radially) is always $>50\%$ at all radii, which validates our interpretation. 
We postpone a detailed investigation of the 3D kinematics of gas in the simulations to a future study, together with the analysis of the mock velocity fields.
However, the considerations above exclude a $100\kpc$-scale galactic fountain as the origin of the flows shown in panels (\emph{a}) and (\emph{d}) of Fig.\,\ref{f:gas_flows}.

For the reasons discussed above, the net flow is likely a more robust quantity to study for our purpose.
The central and rightmost panels of Fig.\,\ref{f:gas_flows} focus on the net flow and show that, in general, there is a tendency for the gas to flow inwards at all radii.
Focussing on the \hi, the radial speed of this inward flow is very similar for the two simulation suites (panel \emph{c}). 
The small amount of \hi\ located at the virial radius $r_{200}$ ($\simeq\!200\kpc$ for the systems studied here) flows in with speeds of $\sim50-100\kms$; this velocity decreases progressively as the gas interacts with the pre-existing CGM, down to about $10\kms$ at $r\simeq10\!-\!30\kpc$ as the material joins the \hi\ disc and proceeds inwards.
However, the median profiles for the \hi\ accretion rates in TNG50 and FIRE-2 galaxies (panel \emph{f}) are quite different, due to the substantially diverse density profiles predicted by the two suites. 
FIRE-2 galaxies feature an \hi\ mass accretion rate of a few $\msunyr$ in all shells with $r\!<\!20\kpc$, down to the galaxy centre. 
The total gas mass accretion rate (panel \emph{e}) in these regions is about twice that provided by \hi\ alone, and with values between $2$ and $10\msunyr$ it closely matches the range of SFRs of FIRE-2 galaxies (Table \ref{t:sample_sims}), supporting a connection between gas accretion and star formation processes.
The accretion proceeds steadily also at $r\!>\!20\kpc$, although it is not traced any longer by the \hi\ phase.
In TNG50 galaxies, instead, $\dot{M}_{\rm HI}$ peaks at $30\!-\!40\kpc$ from the galactic centre, which is approximately the region where the inflowing filaments join the disc in the example galaxy discussed in Section\,\ref{ss:single_galaxy}, and decreases drastically in the innermost regions of the disc.
Considering that the inflow speeds are very similar (panel \emph{c}), this discrepancy is caused by the different density profiles in the TNG50 and FIRE-2 systems, with the latter featuring a much more concentrated gas distribution than the former, likely caused by the lack of AGN feedback in FIRE-2.
The \hi\ accretion rate profiles are somewhat sensitive to the \hi-to-H$_2$ partition scheme: the GK11 recipe leads to a factor $\sim2$ lower \hi\ flow rate in TNG50 galaxies, and an almost null \hi\ accretion in the innermost $4\kpc$ of FIRE-2 galaxies. 
However, the discrepancy between the two simulation suites is visible also in the accretion profiles of the total gas content (panel \emph{e}), which strengthens our interpretation of the gas flow.

Fig.\,\ref{f:gas_lowZ_flows} shows the net mass flow rate for gas with metallicity $<0.5\zsun$.
As expected, the gas that flows in from the outer regions of the halo has a low metal content, but it gets progressively polluted as it approaches the disc.
We notice that the $\dot{M}$ profile for the low-$Z$ gas and the $\dot{M}_{\rm HI}$ profile shown in panel (\emph{f}) of Fig.\,(\ref{f:gas_flows}) are complementary (one increases where the other decreases), which can be due to the pollution promoting the cooling of the inflowing gas with subsequent increase of its \hi\ fraction, a physical process that has been explored in the literature using wind-tunnel simulations at pc-scale resolution \citep{Marinacci+10, Armillotta+17}.
Another possibility, which may coexist with the previous one, is that high-Z \hi\ in the disc is dragged inwards by the accreting flow of low-Z ionised gas from the CGM.
For a more in-depth investigation of gas flows in MW-like  galaxies in the FIRE-2 simulations we refer the reader to \citet{Trapp+22}.

The comparison of panels (\emph{e}) and (\emph{f}) of Fig.\,\ref{f:gas_flows} shows that, in regions sufficiently close to the \hi\ disc ($r\!\lesssim\!30\!-\!50\kpc$), the $\dot{M}_{\rm HI}$ and the $\dot{M}$ profiles are similar (especially in TNG50 galaxies), indicating that the \hi\ is a good tracer of gas inflows.
If correct, this prediction supports the idea that deep \hi\ observations, such as those provided by the MHONGOOSE survey, are key to provide insight into the process of gas accretion onto galaxy discs.
However, the accretion picture shown by the simulations studied here appears to be in tension with recent determination of the distribution of the extraplanar warm ($\sim10^4\!-\!10^5\K$) gas around the Milky Way, where virtually all gas traced by \hi\ and metal low-ions is confined within a height of $\simeq10\kpc$ from the disc \citep{Lehner+22,Choi+24} and is formed, in most parts, by gas participating in the galactic fountain cycle \citep{Marasco+22}.
In other words, there is no observational evidence for gas accretion onto the Milky Way from $T\lesssim10^5\K$ clouds located at several tens of $\kpc$ from the disc.
Assessing the magnitude and the implications of such tension would require a dedicated analysis which goes beyond the scope of this work.

\subsection{Environment}\label{ss:environment}
One may wonder whether the differences between the observed and the simulated galaxy samples can be caused by differences in the local environment.
MHONGOOSE galaxies are in fact specifically selected to be either isolated or to belong to the outer regions of small groups \citepalias{deBlok+24}. 
Target galaxies in the TNG50 MW/M31-like sample, for instance, have no companion with $M_\star\!>\!10^{10.5}\msun$ at distances $<500\kpc$ by construction, but lower mass companions can be present and may have an impact on the \hi\ distribution and kinematics of the main galaxy.

We characterise the local environment of each galaxy in the TNG50 and the observed samples by determining the number $N_{\rm env}$ of satellites with $M_\star\!>\!10^{7.5}\msun$ and $M_{\rm HI}\!>\!10^{7}\msun$ within $250\kpc$ of projected distance and $\pm500\kms$ of line-of-sight velocity from the main galaxy.
For the simulations we analyse the TNG50 sample alone because, unlike FIRE-2, the associated satellite catalogue stores information on the \hi\ content of each sub-halo, which is needed for our analysis.
We stress that $N_{\rm env}$ can be determined in an identical manner in both observations and simulations, with the caveat that satellite stellar masses in the observed sample are determined from \emph{gri} photometry from the DR10 of the Dark Energy Camera Legacy Survey \citep[DECaLS;][]{Dey+19}, where the apparent magnitudes were converted to $M_\star$ assuming the distances of the central galaxies, and using the relations given in \citet{Taylor+11}.

We limit the count of satellites to those having non-negligible \hi\ content because this is how companion galaxies were identified in the data.
An advantage is that their line-of-sight velocity relative to the main galaxy can be measured, without requiring spectroscopic redshift measurements from other facilities.
Given that the lowest \hi\ mass companion presented in \citetalias{deBlok+24} is $\sim10^{7}\msun$ (see their Fig.\,16), we determine $N_{\rm env}$ in TNG50 using this \hi\ mass threshold, which implies that the \hi\ content of the satellite is sampled with at least $100$ gas particles.
We notice that relaxing the requirement on the \hi\ content has a strong impact on our results as, in our TNG50 sample, there are more satellites with $M_{\rm HI}$ below the requested threshold than above.

\begin{figure}
\begin{center}
\includegraphics[width=0.45\textwidth]{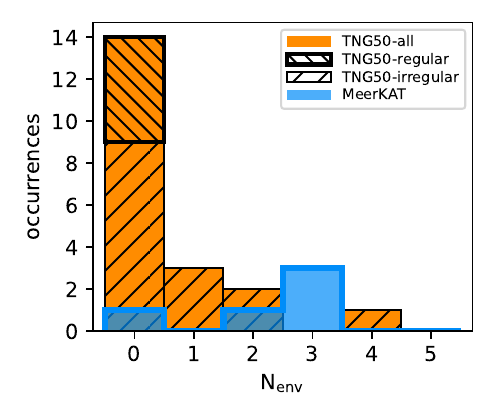}
\caption{Distribution of the local environment estimator $N_{\rm env}$, defined as the number of companions with $M_\star\!>\!10^{7.5}\msun$ and $M_{\rm HI}\!>\!10^7\msun$ detected within $250\kpc$ of projected distance and $\pm500\kms$ of line-of-sight velocity from the main galaxy, for the MeerKAT (blue histogram) and the TNG50 (orange histogram) galaxy samples. The densely and sparsely hatched histograms show the regular and irregular TNG50 populations, respectively (see Fig.\,\ref{f:regular_vs_diffuse_mom0}).}
\label{f:environment}
\end{center}
\end{figure}

\begin{figure}
\begin{center}
\includegraphics[width=0.4\textwidth]{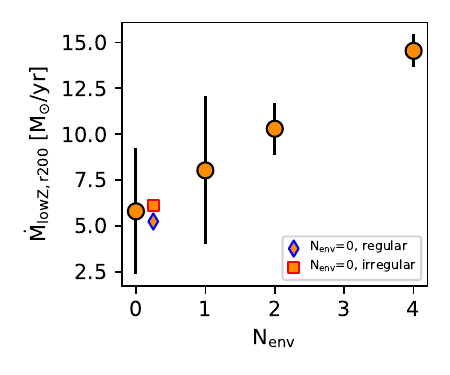}
\caption{Mass accretion rate of metal-poor ($Z\!<\!0.5\zsun$) gas, computed around $r=r_{200}\simeq 200\kpc$, as a function of the environment estimator $N_{\rm env}$, for TNG50 galaxies. Error bars account for the combined effect of galaxy-to-galaxy scatter and radial fluctuations in the $\dot{M}$ estimate around $r_{200}$ in individual systems. The diamond and squared markers show, respectively, the regular and irregular galaxies with $N_{\rm env}\!=\!0$. Galaxies with more satellites show larger accretion rates.}
\label{f:Nsat_vs_accretion}
\end{center} 
\end{figure}

We show the distribution of $N_{\rm env}$ for the two samples in Fig.\,\ref{f:environment}.
There is a substantial difference in the typical environment of the two samples: $14$ out of $20$ TNG50 galaxies have no associated satellite with the properties requested, and only the remaining $6$ appear to have $N_{\rm env}$ compatible with the observations ($N_{\rm env}\!=\!2-3$, with the exception of J0342-47 that has $N_{\rm env}\!=\!0$).
In Fig.\,\ref{f:environment} we have also split the TNG50 sample in two sub-samples, regular (densely-hatched histogram) and irregular (sparsely-hatched histogram), on the basis of the \hi\ morphological classification of Fig.\,\ref{f:regular_vs_diffuse_mom0}.
Interestingly, all simulated galaxies with $N_{\rm env}\!>\!1$ fall in the irregular category, which suggests the existence of a correlation between the presence of nearby companions and that of \hi\ filaments and irregular \hi\ envelopes surrounding galaxy discs.

However, the correlation between satellites and \hi\ features in the galaxy outskirts does not necessarily imply a causal connection between the two. 
In fact, the presence of companions may be caused by an overall more intense accretion flow from the cosmic web, which brings in both satellites and fresh gas from the intergalactic medium (IGM).
We explore this scenario in Fig.\,\ref{f:Nsat_vs_accretion}, where we show the mean mass accretion rate of $Z\!<\!0.5\zsun$ gas, averaged over $150\!<\!r\!<\!250\kpc$ (that is, around the virial radius $r_{200}$), as a function of $N_{\rm env}$. 
The positive correlation between the two quantities indicates that environments with a larger number of satellites, which are those hosting central galaxies with a more irregular \hi\ morphology (Fig.\,\ref{f:environment}), accrete low-$Z$ gas from the IGM at a faster rate. 
This correlation is maintained if we focus on the total accreted gas, rather than on the low-$Z$ component alone. 

Finally, we remark that the number of low-$M_\star$ companions predicted by the simulations depends not only on the accuracy of the model in treating the environment physics, but also on the resulting galaxy stellar mass function. 
In contrast with other simulation suites (including the larger-volume runs from the IllustrisTNG project) which work on larger volumes and are calibrated to reproduce the statistics of the overall galaxy population, TNG50 is not tuned to output the observed galaxy statistics by construction.
Although this does not imply that the resulting satellite statistics are incorrect, caution must be taken when generalising the results presented here.

\subsection{What drives the difference between the regular and the irregular populations?}
\label{ss:regular_vs_irregular_origin}
An important question to address is what is the primary physical mechanism that drives the irregularities in the \hi\ morphology and kinematics of the simulated galaxies.
We have checked that there are no major differences in the $M_\star$, $M_{200}$ or SFR distribution between the regular and irregular galaxies, indicating that total dynamical mass or stellar feedback do not drive the difference between the two populations.
Based on the analysis of Section \ref{ss:environment}, we can certainly assert that - at least in TNG50 - the environment is one of the factors that regulate the presence of low-$N_{\rm HI}$ features at the periphery of galaxy discs, which is frequent in the simulated systems but is not visible in the observed ones: higher mass accretion from the cosmic web is typically associated with more companions that stream inwards with the accretion flow, producing a diffuse gas component around the galaxy which is expected to be observed with deep \hi\ observations.

To verify that higher accretion rates onto halos correspond to stronger gas inflows onto galaxy discs \citep[which may not necessarily be the case, see][]{vandeVoort+11}, we have inspected the radial profiles of the net \hi\ flow (see panel \emph{e} in Fig.\,\ref{f:gas_flows}) for the regular and irregular TNG50 systems separately. 
We found a peak accretion value at $20\!<R\!<\!30\kpc$ of $\sim2.5\msunyr$ for the former and of $\sim5\msunyr$ for the latter.
FIRE-2 galaxies behave similarly, although the largest difference in accretion rates between the regular and irregular systems is visible at $R\!<\!3\kpc$, where we infer $\sim3\msunyr$ for the former and $\sim10\msunyr$ for the latter.
This confirms that the large-scale environment has an impact on the gas accretion onto galaxies, at least in the simulations studied here.
We notice that the TNG50 irregular population hosts a higher fraction of systems in the so-called `thermal' AGN-feedback mode (6/15, compared to 1/5 for the regular ones). 
This is a direct consequence of the AGN feedback implementation scheme 
as these galaxies host SMBHs with masses just below $10^8\msun$,
a threshold where the $z\!=\!0$ SMBH population has a clear change in accretion rate \citep{Weinberger+17}.
These systems, marked by a red dot in Fig\,\ref{f:regular_vs_diffuse_mom0}, do not seem to have a distinct morphology or kinematics.

We believe that the fact that the observed galaxies feature regular and smooth \hi\ discs with little diffuse low-N$_{HI}$ gas - in spite of inhabiting environments that are typically richer than those of most simulated galaxies (Fig.\,\ref{f:environment}) - indicates that a different, gentler, disc-halo interplay is at work in the observed Universe.
More specifically, the processes of inflows and outflows that regulate the evolution of nearby late-type galaxies occurs without significantly disturbing the morphology and kinematics of \hi\ discs, which is something that the simulations analysed here struggle to achieve.
A similar conclusion was reached by \citet{Marasco+23a} by studying the galactic wind properties in a sample of $19$ nearby low-$M_\star$ starbursts.
The fact that both regularly star forming galaxies and starburst systems feature similarly mild gas flows around their discs is indicative of an overall gas circulation in nearby galaxies that is gentler than that predicted by the simulations.
In this context, the galactic fountain framework proposed by \citet{Fraternali17}, where the mild gas circulation produced by stellar feedback drives the cooling of the lower layers of the corona, seems to be a very promising model to interpret the observations.

\subsection{The relation between moment-2 and stellar feedback}\label{ss:mom2_feedback}
\begin{figure*}
\begin{center}
\includegraphics[width=\textwidth]{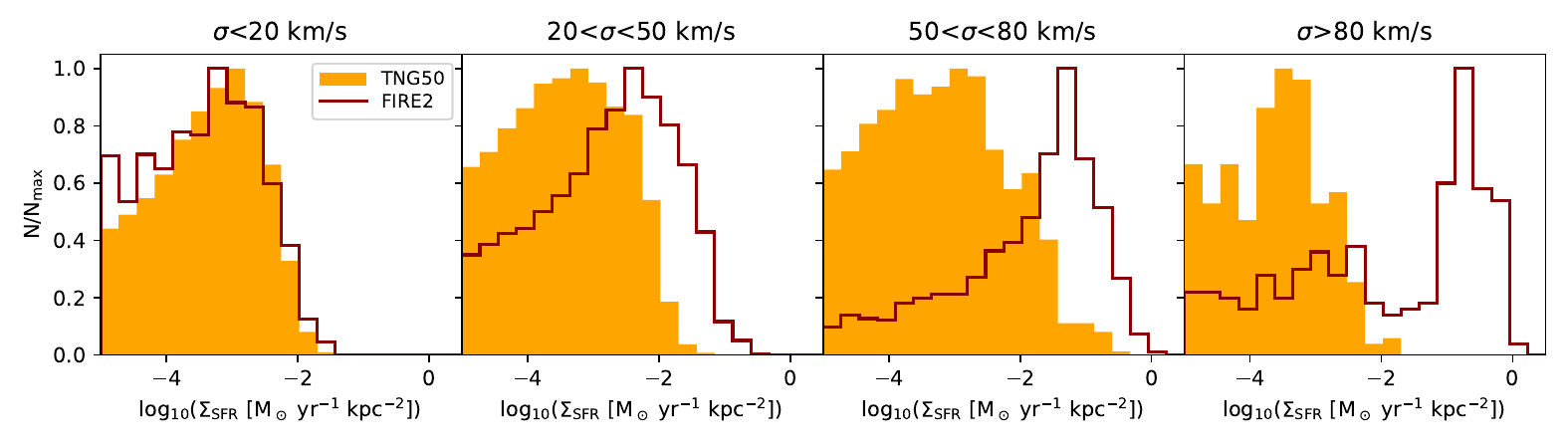}
\caption{SFR surface density distribution in TNG50 (orange-filled histograms) and FIRE-2 (dark red histograms) galaxies, for the HR case. Each panel accounts for spaxels in a given range of \hi\ velocity dispersion, indicated on top.}
\label{f:SFRD_vs_mom2}
\end{center}
\end{figure*}
In Sections \ref{ss:single_galaxy} and \ref{ss:mom0mom2} we have shown that simulated galaxies feature several high-velocity peaks in their moment-2 maps, which lead to a high-velocity tail in the $\sigma$ distribution that is not present in the observations (right panel of Fig.\,\ref{f:mom0mom2_1D}).

There are various possible origins for the high-$\sigma$ \hi\ regions.
Deep gravitational potential wells caused by massive bulges can promote turbulence within the ISM due to the increased shear in the innermost regions \citep[e.g.][]{Martig+09}, but this effect is typically limited to the central kpc \citep{Gensior+20} thus cannot be resolved by our mock \hi\ data, not even for the HR case.
A more promising option is that these features are due to the effect of stellar feedback, which injects both momentum and thermal energy into the gas surrounding star-forming regions\footnote{Detailed descriptions of stellar feedback recipes for TNG50 and FIRE-2 simulations are provided by \citet{Pillepich+18} and \citet{Hopkins+18b}, respectively.}, leading to galaxy-scale winds and enhancing the velocity dispersion in localised regions of the disc. 
We test this scenario in Fig.\,\ref{f:SFRD_vs_mom2}, which shows histograms of the local SFR surface density ($\Sigma_{\rm SFR}$) in TNG50 and FIRE-2 galaxies, divided into bins of different $\sigma$ as determined from the moment-2 HR maps. 
The $\Sigma_{\rm SFR}$ is extracted from maps that are built using gas particle-based SFRs, assuming the same observational properties (resolution, field of view, galaxy inclination and position angle) adopted for generating the synthetic \hi\ HR data.

Fig.\,\ref{f:SFRD_vs_mom2} shows a clear distinction between FIRE-2 and TNG50 galaxies.
In FIRE-2, the peak of the $\Sigma_{\rm SFR}$ distribution shifts towards progressively higher values for larger $\sigma$ bins, strongly supporting a stellar feedback origin for the high-$\sigma$ \hi\ regions.
TNG50 galaxies show a more complex behaviour: the $\Sigma_{\rm SFR}$ distribution peaks always at a similar value ($\sim10^{-3} \msunyr\kpc^{-1}$) but becomes wider as $\sigma$ increases, with the exception of the highest $\sigma$ bin (rightmost panel of Fig.\,\ref{f:SFRD_vs_mom2}).
This suggests that, in TNG galaxies, star formation is not the primary driver of the high-$\sigma$ \hi\ features.
Projection effects, accretion of pristine gas from the cosmic web, gas recycling, tidal interactions or wet minor mergers can all be responsible for a high-$\sigma$ component that is uncorrelated with the $\Sigma_{\rm SFR}$. 

In Section\,\ref{ss:single_galaxy} we have already discussed a case where high-$\sigma$ peaks appear to be located at the intersection between the \hi\ disc and the surrounding filaments, supporting the existence of a causal connection between gas inflow and velocity dispersion enhancement.
The inspection of the moment-2 maps shown in Figure \ref{f:regular_vs_diffuse_mom2} reveals that there is a difference in the spatial distribution of the velocity dispersion between the two simulated samples: TNG50 systems exhibit multiple $\sigma$ peaks, several of which are located in the outer regions of the discs, while FIRE-2 galaxies show more axisymmetric moment-2 maps, with the highest values that are systematically concentrated in the innermost regions of their discs. 
While the maximum moment-2 values in FIRE-2 galaxies can be found at the galactic centre and are due to beam-smearing, $\sigma$ typically remains above $40\kms$ up to $\sim70"$ from the centre, i.e., a distance much larger than the spatial resolution of the HR case.
The difference between the two simulations can be interpreted in the light of the different gas accretion profiles shown by the TNG50 and FIRE-2 galaxies (Section\,\ref{ss:inflow_outflow}).
In FIRE-2 galaxies, gas streams inwards from the virial radius down to the innermost kpc with an approximately constant rate (bottom-left panel of Fig.\,\ref{f:gas_flows}). 
This promotes an intense star formation activity in the central region of the galaxy, producing the high moment-2 values observed.
It is possible that the lack of AGN feedback in the FIRE-2 runs plays a crucial role in this process, as one would expect that part of the accretion flow feeds the central supermassive black hole, whose feedback would act as a regulator of the flow itself.
In TNG50, the gas accretion rate becomes negligible at $\sim20\kpc$ from the centre, presumably where the inflowing filaments join the \hi\ disc.
This promotes the onset of turbulence also in the outer regions of the galaxy where star formation is negligible, explaining the lack of trends between SFR density and velocity dispersion (Fig.\,\ref{f:SFRD_vs_mom2}). These considerations aside, it is also possible that the different behaviour of TNG50 and FIRE-2 galaxies stems from the different response of the \hi\ to feedback from star formation, due to the different resolution and feedback implementation of the two simulation suites.

\subsection{The complexity of the \hi\ line profiles}\label{ss:line_profiles}
\begin{figure*}
\begin{center}
\includegraphics[width=0.9\textwidth,trim={8.0cm 1.5cm 10.0cm 1.5cm},clip]{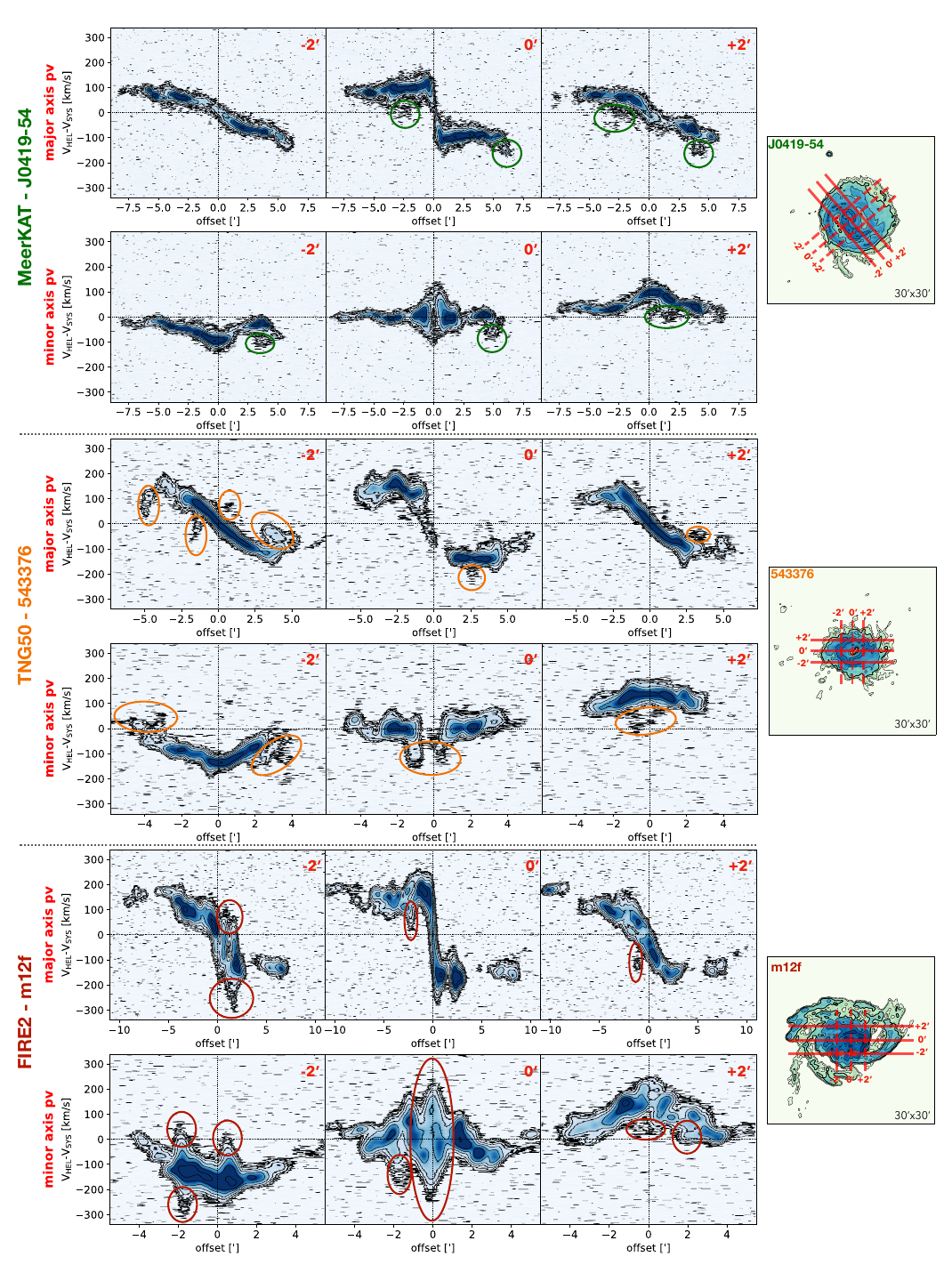}
\caption{\hi\ position-velocity diagrams on slices parallel to the kinematic major and minor axes (first and second rows of each panel set) for J0419-54 (top panels), the TNG50 system 543376 (central panels), and the FIRE-2 system m12f (bottom panels), for the HR case. All panels show the same velocity range on the y-axis. Slice thickness is $4$ pixels ($20"$). Slice offsets are indicated on the top-right corner of each panel. Contours are spaced by powers of $2$, the outermost being at an intensity level of $0.3\miJyb$ ($2\sigma_{\rm noise}$). A negative contour ($-2\sigma_{\rm noise}$) is shown as in grey. Circles highlight the anomalous kinematic features. The rightmost insets show the total \hi\ maps with the cuts parallel to the major and minor axes (solid and dashed lines, respectively) used in the pv-slices.} 
\label{f:pv}
\end{center}
\end{figure*}

\begin{figure}
\begin{center}
\includegraphics[width=0.5\textwidth]{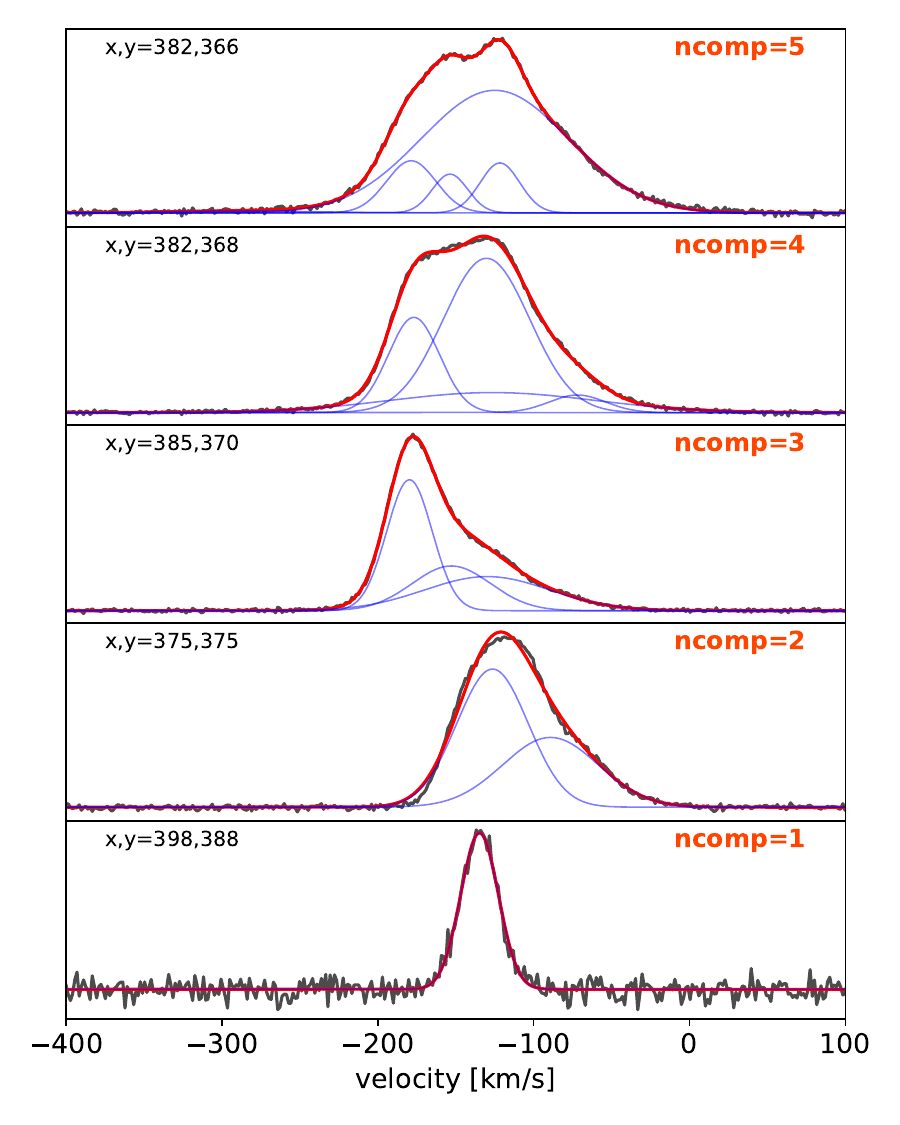}
\caption{Representative \hi\ spectra for the FIRE-2 galaxy m12f (HR), extracted at the pixel positions indicated on the top-left corner of each panel. The grey lines show the mock \hi\ spectra, the red lines show the best-fit multi-Gaussian models, the blue lines show the individual Gaussian components. The number of components used is indicated in the top-right corner of each panel.}
\label{f:gaumodel_test}
\end{center}
\end{figure}

\begin{figure*}
\begin{center}
\includegraphics[width=\textwidth]{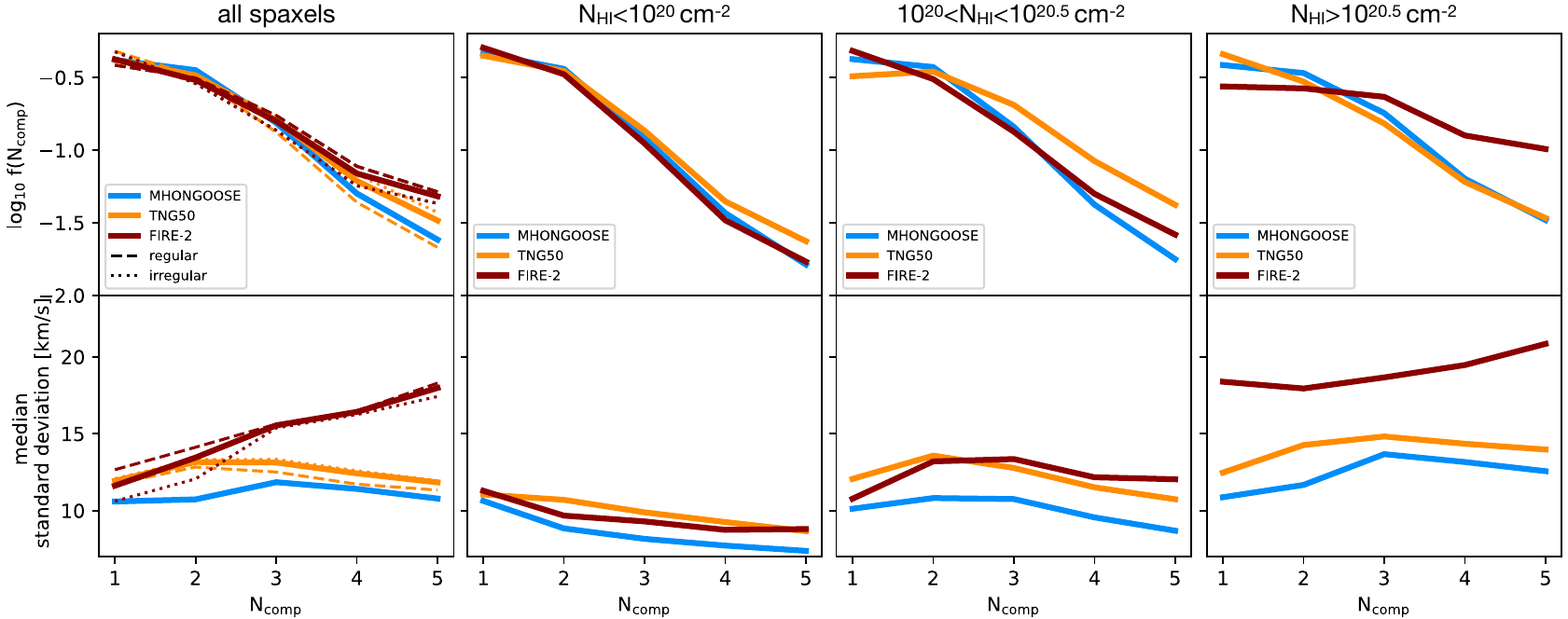}
\caption{Results of our multi-Gaussian fitting for the TNG50 (orange), FIRE-2 (dark red) and MeerKAT (blue) galaxies. \emph{Top panels}: fraction of spectra modelled with $N$ components ($1\!\leq\!N\!\leq\!5$). \emph{Bottom panels:} median value of the distribution of the Gaussian standard deviations computed for all line profiles modelled with $N$ components, as a function of $N$. From left to right, the different panels show our results for spaxels in different ranges of (inclination-corrected) \hi\ column density, indicated on the top of each panel.
The dashed and the dotted curves on the leftmost panels show, respectively, the regular and irregular populations of simulated galaxies.}
\label{f:frac_sigma_ncomp}
\end{center}
\end{figure*}

We now attempt to characterise the typical shape of the \hi\ line profiles in the simulated and the observed systems, with the goal of providing quantitative measurements for their complexity.
This exercise is also useful to understand whether the higher moment-2 values observed in the mocks are due to the presence of multiple components along the line of sight, to broader individual \hi\ components, or to a combination of these effects.

We first provide an example of the different kinematic complexity exhibited by three systems in our samples.
Fig.\,\ref{f:pv} shows a series of position-velocity (pv) slices for the MHONGOOSE galaxy J0419-54 (top panels), the TNG50 galaxy 543376 (central panels), and the FIRE-2 galaxy m12f (bottom panels).
In general the pv slices of the observed and simulated galaxies appear very similar, and are dominated by the presence of a large rotating disc with an approximately flat rotation curve. 
Both in the observations and in the mocks there are very few \hi\ clouds that are kinematically detached from the main disc, and virtually none of these reside in the counter-rotating regions of the pv slices.
These considerations further stress the degree of realism achieved by present-day hydrodynamical simulations of galaxy formation.

Upon closer inspection, Fig.\,\ref{f:pv} reveals two important differences between the MeerKAT observations and the mock data.
First, the pv slices of J0419-54 are `thinner' (in the vertical direction) than those of the other galaxies, and especially than those of m12f.
We have verified that the thermal broadening adopted for the mocks ($\sigma_{\rm th}\!=\!8\kms$) is not the main source of the discrepancy, as models with $\sigma_{\rm th}\!=\!5\kms$ appear almost identical.
To test whether the discrepancy arises from projection effects due to extremely thick discs in the simulations, which would smear the velocity profiles due to the line-of-sight intersecting layers at very different radii, we have determined the \hi\ scale-height radial profiles of the simulated galaxies by fitting their vertical \hi\ distributions in different annuli with a Gaussian function \citep[e.g.][]{Olling95}.
The resulting scale-height profiles show clear signatures of inside-out flaring, with typical values of $0.5\!-\!1\kpc$ at $R\!=\!10\kpc$, in line with the results of \citet{Gensior+23} for simulated MW-like galaxies from the FIREbox suite \citep{Feldmann+23} and compatible with (or only marginally larger than) those inferred in nearby galaxies of similar $M_\star$ \citep[e.g.][]{OBrien+10,Bacchini+19}, including the Milky Way \citep{Marasco+17}.
Thus we exclude that the thickness of \hi\ discs plays a significative role in the cases studied here.
The second difference is that the simulated galaxies show several \hi\ features located at velocities that are anomalous with respect to those occupied by the bulk of the \hi\ distribution at that specific location.
Following a terminology that has often been adopted in the literature \citep[e.g.][]{FB06, Sancisi+08, Gentile+13, Marasco+19b}, we refer to these features as `extraplanar'.
These are highlighted with circles in Fig.\,\ref{f:pv}, and are more prominent in the simulated galaxies than in J0419-54.
This comparison already indicates that \hi\ line profiles in TNG50 and (especially) FIRE-2 galaxies are individually broader and more complex than those in the observed systems.

In order to quantify the discrepancy between simulations and observations, we perform a multi-Gaussian decomposition of all line profiles within the SoFiA masks.
Our approach consists of fitting a variable number of Gaussian components, ranging from $1$ to $5$, to all profiles whose peak signal-to-noise ratio is above $2$.
The \textsc{find\_peaks} subroutine of the \textsc{scipy} package \citep{scipy} in \textsc{python} is used to identify the locations of multiple peaks, which, if detected, are used as first guess to the fitting procedure.
Following \citet{Marasco+20} and \citet{Marasco+23a}, the optimal number of components is decided locally on the basis of a Kolgomorov–Smirnov (KS) test on the residuals of the fits determined using $N$ and $N+1$ components: if the two distributions of residuals are statistically different, the $N+1$ component model is preferred.
Thus, this method is sensitive to the relative improvements in the data fitting due to the use of increasingly complex model profiles, but it is independent of the `goodness' of the fit in absolute terms.
A `sensitivity threshold' for the KS $p$-value must be adopted to reject the null hypothesis of identical residuals: this value ranges from $0$ (the test is insensitive to all variations in the residuals, hence only one component is needed) to $1$ (the test is overly-sensitive and all $5$ components are always needed).
Visual inspection of the resulting model profiles overlaid on top of the data indicate that a threshold very close to unity (1-$\epsilon$, with $\epsilon\!=\!5\times10^{-5}$) is required in order to model the most complex line shapes.

We demonstrate our fitting procedure in Fig.\,\ref{f:gaumodel_test}, which presents a collection of five representative \hi\ line profiles of the FIRE-2 galaxy m12f (HR), showing separately the mock spectra (in grey), the multi-Gaussian fits (in red) and the single Gaussian (in blue).
All of these velocity profiles are well sampled by the simulation, as the number of gas particles with \hi\ fraction $>10^{-5}$ that contribute to the \hi\ lines ranges between a few thousands to a few tens of thousands.
The various panels of Fig.\,\ref{f:gaumodel_test} demonstrate that the multi-Gaussian model fits the mock data very well, using a variable number of components depending on the complexity of the line profile analysed. 
We notice that, even with the relatively large sensitivity threshold adopted, the approach taken prefers a limited number of components and does not `overfit' the mock. 
This is particularly visible in the fourth panel of Fig.\,\ref{f:gaumodel_test}, which gives the impression that the number of components adopted to model the simulated line profile (two) is somewhat conservative.
This indicates that our method of decomposition likely provides a lower limit on the number of components.

We use this approach to produce multi-Gaussian fits for all TNG50, FIRE-2 and MeerKAT systems.
To compare our results amongst the three samples, we make use of two quantities. 
The first is the fraction of spectra modelled with a given number $N_{\rm comp}$ of components. 
The second quantity is computed by building, using all spectra modelled with $N_{\rm comp}$ components, the distribution of the Gaussian standard deviations associated with each component, and then extracting the median of such distribution.
In practice, the first quantity accounts for possible differences in the typical number of \hi\ components, whereas the second is informative of the typical broadening of the individual components. 
Both quantities are computed using only the unmasked spectra of all galaxies of a given sample, that is, using only the spectra that contribute to the various moment maps.
The leftmost panels of Fig.\,\ref{f:frac_sigma_ncomp} show how these two quantities vary as a function of the number of components in the three galaxy samples, and for the regular and irregular populations separately.
Clearly, the origin of the discrepancy between the simulated and the observed galaxies is twofold: with respect to the mocks, the \hi\ profiles in the real galaxies are characterised by a typically lower number of components, each having on average lower standard deviations.
In particular, FIRE-2 galaxies are those that show the largest deviations from the observations, featuring a factor $\sim2$ more 5-component profiles than the MeerKAT galaxies, and a factor of $\sim1.7$ higher standard deviations at $N\!=\!5$. 
As expected, the regular population provides a marginally better match to the data.

To identify whether the trends visible in the leftmost panels Fig.\,\ref{f:frac_sigma_ncomp} depend on the \hi\ column density, we have repeated our analysis using only spaxels in given ranges of (inclination-corrected) $N_{\rm HI}$: below $10^{20}\cmmq$, between $10^{20}$ and $10^{20.5}\cmmq$, and above $10^{20.5}\cmmq$.
These are shown in the second, third and fourth columns of Fig.\,\ref{f:frac_sigma_ncomp}.
Unsurprisingly, we find a larger number of components and a stronger broadening for increasing values of $N_{\rm HI}$.
Interestingly, the largest deviations between the mocks and the data are visible for the highest $N_{\rm HI}$ bin in FIRE-2, and for the intermediate $N_{\rm HI}$ bin for TNG50.
This strengthens our idea that the peculiar kinematics of FIRE-2 galaxies is driven primarily by stellar feedback, which  affects preferentially the high-N$_{\rm HI}$ gas, whereas the high-$\sigma$ regions in TNG50 galaxies are produced by a different mechanism such as accretion of cold gas filaments at the periphery of their disc, as discussed in Section \ref{ss:mom2_feedback}.

We can robustly conclude that the simulated galaxies, and in particular those from the FIRE-2 suite, feature more complex and broader \hi\ line profiles than the observed galaxies, and that such peculiar kinematics is caused by a more complex gas cycle within and around their discs.

\section{Conclusions}\label{s:conclusions}
The distribution and kinematics of gas in and around galaxy discs is influenced by both internal and external processes such as stellar and AGN feedback and gas accretion from the circumgalactic medium.
These processes produce a large-scale gas circulation whose impact is expected to be more visible in the low column density gas located at the interface region between the galactic disc and the halo.
Deep observations are needed to probe gas in this region, and state-of-the-art galaxy evolution models are required for their interpretation.

In this study we have used wide-field ($1\de\times1\de$), spatially resolved (down to $22''$), high-sensitivity ($\sim10^{18}\cmmq$) \hi\ observations with the MeerKAT radio telescope of a sample of five Milky Way-like galaxies. 
The typical galaxy distance of $20\Mpc$ allows us to sample a region of few hundred kpc around their disc with a spatial resolution of $2\!-\!6$ kpc, depending on the angular resolution adopted.
We have compared these data with synthetic \hi\ observations generated from a sample of $25$ simulated star-forming galaxies from the TNG50 ($20$) and FIRE-2 ($5$) suites of cosmological hydrodynamical simulations in the $\Lambda$CDM framework.
The simulated galaxies are selected to be similar to the observed ones in terms of stellar mass, SFR and \hi\ content, and the mock \hi\ datacubes (and associated moment maps) are built using the same sensitivity, resolution and field of view as the MeerKAT observations.
This approach allows us to process both observations and simulations with a similar methodology, limiting the bias in the comparison between the two.
Our results can be summarised as follows.
\begin{itemize}
    \item By exploring different recipes \citep{BlitzRosolowsky06, GnedinKravtsov11, Krumholz13} for the atomic-to-molecular hydrogen partitioning in the simulations, we found that the \citet{BlitzRosolowsky06} approach leads to H$_2$-to-\hi\ mass ratios that are in good agreement with those inferred in the local Universe (Fig.\,\ref{f:MH2_over_MHI}). The other recipes overestimate the H$_2$-to-\hi\ mass ratio by up to an order of magnitude, and we advise against their use to infer \hi\ and H$_2$ masses in TNG50 and FIRE-2 galaxies.

    \item Compared against the MeerKAT sample, the simulated galaxies show on average an excess of low column density gas ($N_{\rm HI}\!<\!10^{20}\cmmq$) in their outskirts (left panels of Fig.\,\ref{f:mom0mom2_1D} and Fig.\,\ref{f:fHI_faint_LR}), with several systems having $10\!-\!20\%$ of their total \hi\ content distributed at $50\!-\!150\kpc$ from their disc (Fig.\,\ref{f:cumulative_HR}). 
    
    \item While all galaxies in the observed sample appear as regular discs, the simulated galaxies frequently show disturbed and irregular \hi\ morphologies. In particular, all systems whose \hi\ spatial distribution is dominated by column densities $\!<\!10^{20}\cmmq$ show strong irregularities (Fig.\,\ref{f:regular_vs_diffuse_mom0}).
    
    \item The local environment plays a key role in shaping the \hi\ distribution of simulated galaxies. \hi-irregular systems inhabit halos with more satellites (Fig.\,\ref{f:environment}) and with higher gas accretion rates (Fig.\,\ref{f:Nsat_vs_accretion}) than those hosting more regular \hi\ discs. However, the observed galaxies always show regular \hi\ discs in spite of hosting a number of \hi-rich satellites typically larger than the satellites found near simulated systems.

    \item The simulated galaxies show a high-velocity tail in their moment-2 distribution which is not visible in the real systems (right panels of Fig.\,\ref{f:mom0mom2_1D}). This tail is produced by line profiles that have, on average, a broader and overall more complex shape than those observed with MeerKAT (Figs.\,\ref{f:pv} and \ref{f:frac_sigma_ncomp}). These features are stronger in FIRE-2 than in TNG50.    
\end{itemize}

The complex kinematics shown by the simulated galaxies is caused by the combined effect of extragalactic gas accretion and stellar feedback. 
In FIRE-2 galaxies, gas streams inwards from the virial radius down to the innermost kiloparsec at an approximately constant rate, promoting an intense star formation activity (Section \ref{ss:inflow_outflow}) that leads to the high moment-2 values concentrated towards the central regions of the galaxy (Fig.\,\ref{f:regular_vs_diffuse_mom2}). In TNG50, the inwards accretion flow stops at the disc periphery, promoting turbulence also in the outer regions of the galaxy, where star formation is negligible.
Real galaxies show more quiet kinematics. 
Our results support a scenario where gas circulation caused by stellar feedback and accretion from the cosmic web in real galaxies is gentler than in the simulated ones. 
It leads to \hi\ discs that are less turbulent, feature less diffuse, low-$N_{\rm HI}$ ($<\!10^{20}\cmmq$) emission, and overall show a more regular morphology than those in the cosmological hydrodynamical simulations considered here.
A promising scenario that could explain the observed \hi\ properties of nearby disc galaxies is that of the galactic fountain \citep{Fraternali17}.

In spite of their very different characteristics (hydro code, mass resolution and sub-grid physics), both the TNG50 and the FIRE-2 simulations have shown shortcomings when compared to the data.
One may wonder whether the discrepancies between the simulated and the observed galaxies are due to some missing physics in the models (e.g. radiative transfer, lack of resolution in the CGM, cosmic rays) or are intrinsic to all galaxy evolutionary models in the $\Lambda$CDM framework.
Although this is a question for future studies, we argue that the method developed in this work, based on a detailed comparison between the \hi\ properties in simulated and observed galaxy discs, is an ideal tool to provide the answer.

\begin{acknowledgements}
The authors thank an anonymous referee for a thoughtful and constructive report.
AM acknowledges funding from the INAF Mini Grant 2023 program
`The quest for gas accretion: modelling the dynamics of extra-planar gas in nearby galaxies'.
AM thanks Adam R.\,H. Stevens for providing the code to implement the various recipes for the molecular-to-atomic hydrogen separation in FIRE-2 galaxies.
The MeerKAT telescope is operated by the South African Radio Astronomy Observatory, which is a facility
of the National Research Foundation, an agency of the Department of Science and Innovation.
This work has received funding from the European Research Council
(ERC) under the European Union’s Horizon 2020 research and innovation
programme (grant agreement No. 882793 ``MeerGas'').
This work was funded by the European Union (ERC, FLOWS, 101096087). Views and opinions expressed are however those of the author(s) only and do not necessarily reflect those of the European Union or the European Research Council. Neither the European Union nor the granting authority can be held responsible for them.
PK is partially supported by the BMBF project 05A23PC1 for D-MeerKAT.
LC acknowledges funding from the Chilean Agencia Nacional de Investigación y Desarrollo (ANID) through Fondo Nacional de Desarrollo Científico y Tecnologico (FONDECYT) Regular Project 1210992. 
KAO acknowledges support by the Royal Society through a Dorothy Hodgkin Fellowship (DHF/R1/231105). 
\end{acknowledgements}


\bibliographystyle{aa} 
\bibliography{MHONGOOSE_sims} 

\begin{appendix}
\onecolumn
\section{Additional material} \label{a:supplementary}
\begin{figure*}[h!]
\begin{center}
\includegraphics[width=0.83\textwidth,trim={0 4.0cm 2.0cm 0},clip]{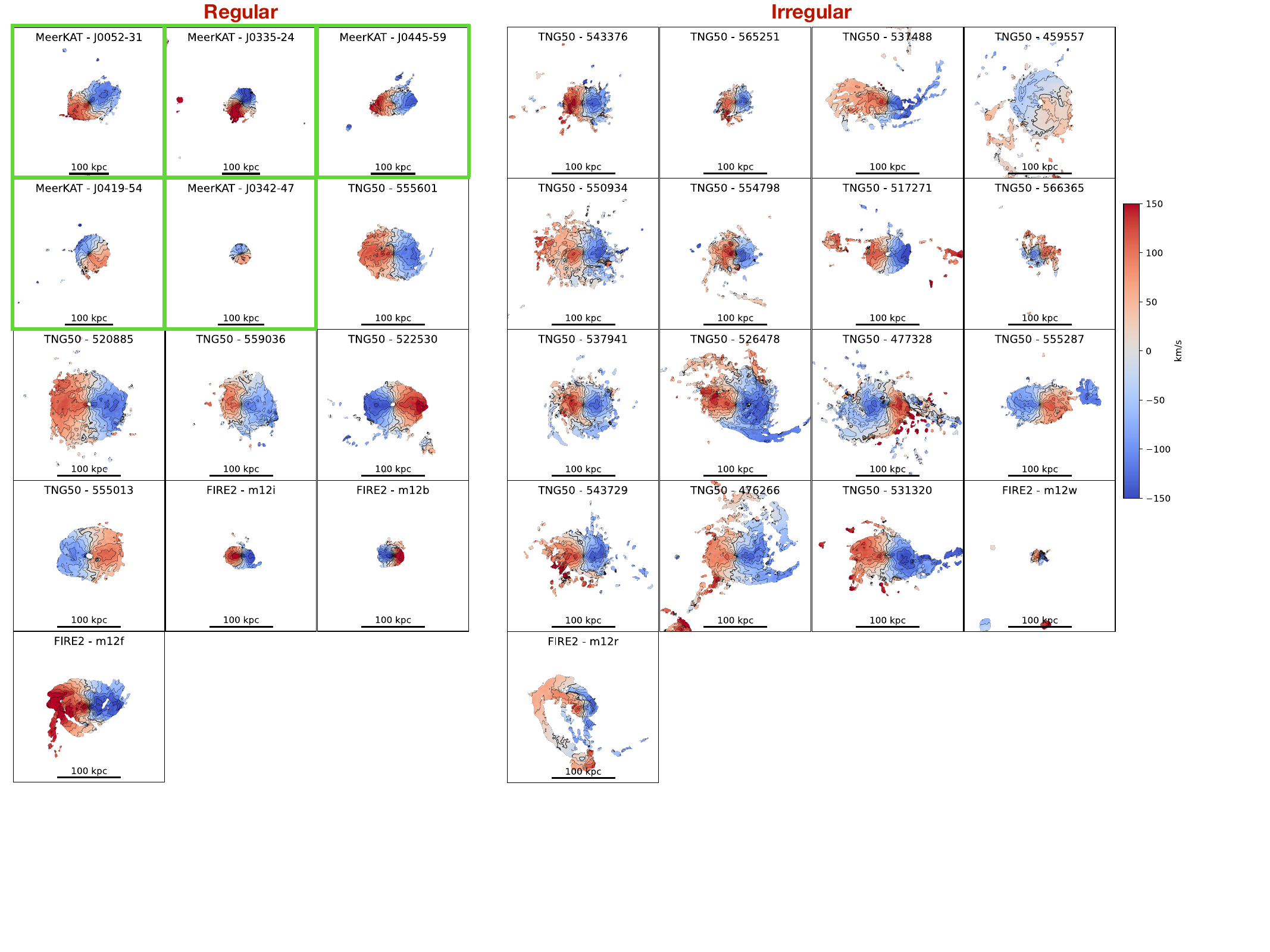}
\caption{As in Fig.\,\ref{f:regular_vs_diffuse_mom0}, but for the moment-1 HR maps. The thicker contour shows the systemic velocity. Subsequent contours are spaced by $25\kms$.}
\label{f:regular_vs_diffuse_mom1}
\end{center}
\end{figure*}

\begin{figure*}[h!]
\begin{center}
\includegraphics[width=0.83\textwidth,trim={0 4.0cm 2.0cm 0},clip]{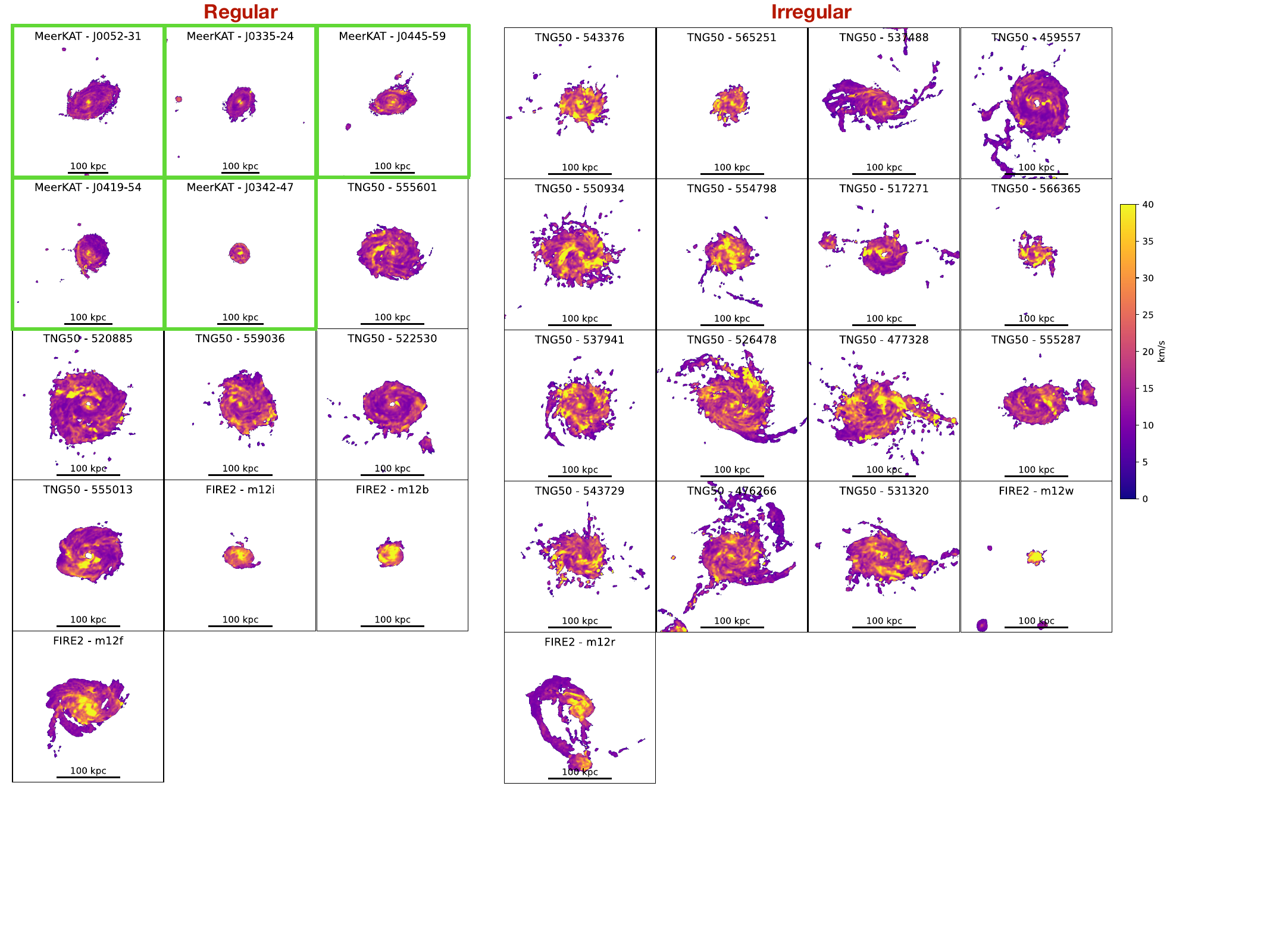}
\caption{As in Fig.\,\ref{f:regular_vs_diffuse_mom0}, but for the moment-2 HR maps.}
\label{f:regular_vs_diffuse_mom2}
\end{center}
\end{figure*}

\begin{figure*}[h!]
\begin{center}
\includegraphics[width=0.73\textwidth]{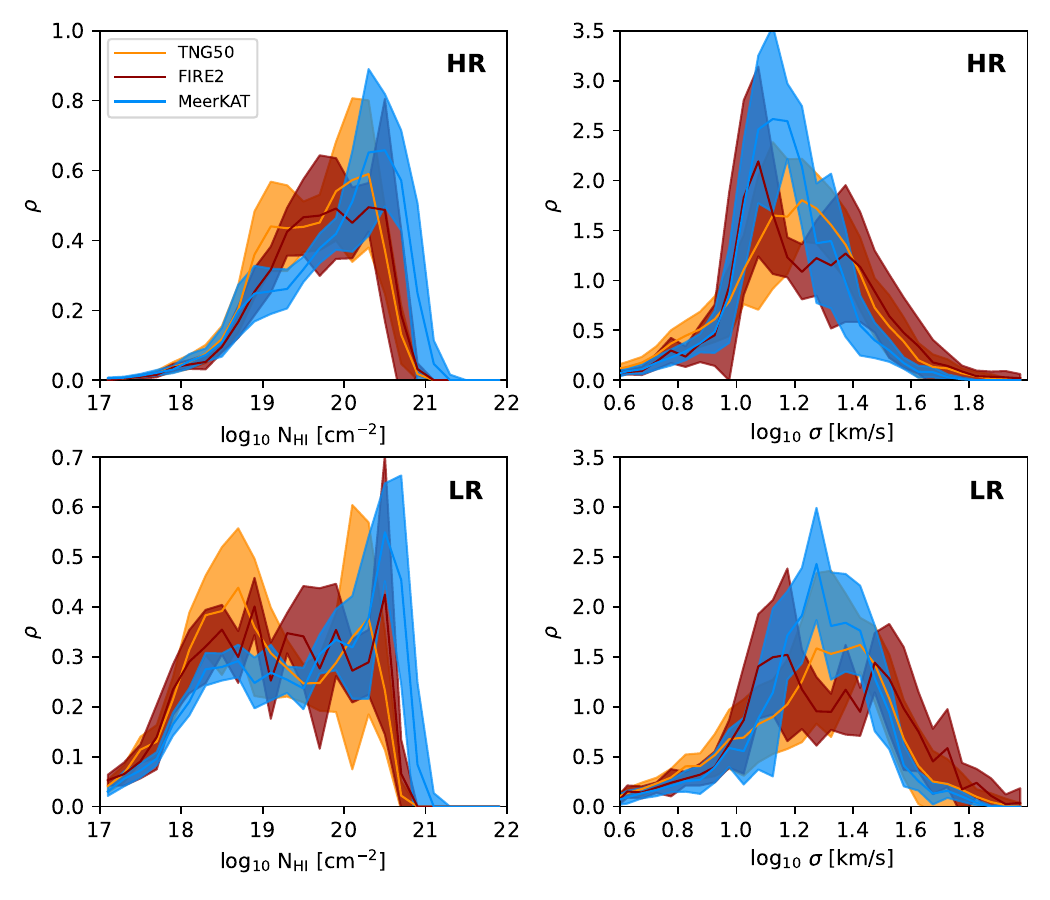}
\caption{As in Fig.\,\ref{f:mom0mom2_1D}, but using the H$_2$-to-\hi\ partition recipe of \citet{GnedinKravtsov11}. Compared to the \citet{BlitzRosolowsky06} prescription, the column density range spanned by the FIRE-2 galaxies is compressed to lower values and the tail at $N_{\rm HI}\!>\!10^{21}\cmmq$ has disappeared.}
\label{f:mom0mom2_1D_MH2GK}
\end{center}
\end{figure*}
\end{appendix}

\end{document}